\documentclass[%
 reprint,
amsmath,amssymb,
pra,
]{revtex4-2}

\usepackage{graphicx}
\usepackage{dcolumn}
\usepackage{bm}
\usepackage[mathlines]{lineno}

\usepackage{amsmath}
\usepackage{amsfonts}
\usepackage{mathtools}

\usepackage{hyperref}
\hypersetup{
  colorlinks=true,
  linkcolor=red,
  citecolor=blue,
}
\usepackage{cleveref}

\begin{document}

\title{Identifying robust features of community structure in complex networks}

\author{Karsten~N.~Economou}
\email{KarstenEconomou@dal.ca}
\author{Cassie~R.~Norman}
\author{Wendy~C.~Gentleman}
\affiliation{
 Department of Engineering Mathematics and Internetworking, Dalhousie University, Halifax, Nova Scotia, Canada
}
\date{\today}

\begin{abstract}
Network science has presented community detection as a valuable tool for revealing functional modules in complex systems rooted in the wiring architectures of complex networks.
The varying procedures of community detection can produce, however, divisions of a network into communities that vary considerably in structure but are deemed to be of similar merit.
This is especially problematic when the network is constructed on uncertain data, since small changes to the network's configuration can cause radically different structure to be detected.
To reconcile with the ambiguity in interpreting degenerate network partitions as representations of the underlying system function, we introduce a recursive significance clustering scheme that identifies the subsets of nodes having stable joint community assignments under network perturbation.
These robust node groups are referred to here as cores, and represent well-supported features of network structure as distinct from the nodes with unstable community assignments.
We show that cores characterize the variability inherent to non-overlapping community structure in networks and are cohesive under temporal evolution of the network.
\end{abstract}

\maketitle

\section{\label{sec:introduction} Introduction}
Characterizing complex systems is one of the great modern scientific challenges due to the difficulty of accurately and completely deriving their collective behaviors from their atomic components \cite{Strogatz2001}.
Enter complex networks, whose anatomies encode the intricate interactions between the discrete components of a complex system \cite{AlbertBarabasi2002,Newman2003}.
Despite varying greatly in form, composition, and scope, complex networks often share common organizing principles \cite{Barabasi2016}.

Of particular interest in network science is the structure of a network's wiring architecture, since
structure of the network represents function of the system \cite{Strogatz2001}.
One manner in which many complex networks arrange themselves is into \emph{communities}: groups of nodes with high internal link density that are sparsely linked with others \cite{GirvanNewman2002}.
In non-overlapping community structure, each node is assigned to exactly one community.
Communities may be understood as representations of functional modules embedded within the networked system \cite{Fortunato2010}.
Scaling the lens of analysis to the mesoscopic interactions between communities omits superfluous detail without obscuring significant relationships in the system, providing a simplified view of interaction patterns \cite{RosvallBergstrom2008}.

As community detection is an NP-hard problem \cite{Brandes2008}, many algorithms utilize stochastic searches \cite{Blondel2008,Waltman2013,Edler2017,Traag2019} to optimize an objective function that idiosyncratically defines the quality of a network partition \cite{Newman2004,Rosvall2009}.
This exposes results to degenerate solutions of dissimilar community structures that are of similar merit \cite{Good2010,Lancichinetti2012,Coscia2014,Rossetti2018,Tandon2019}.
Small changes to the network's components (due to, e.g., measurement noise, model uncertainty, or temporal evolution) can exacerbate this issue, causing radically different structures to be detected \cite{Rossetti2018,Calatayud2019,Nathe2022}.

\begin{figure}
    \centering
    \includegraphics{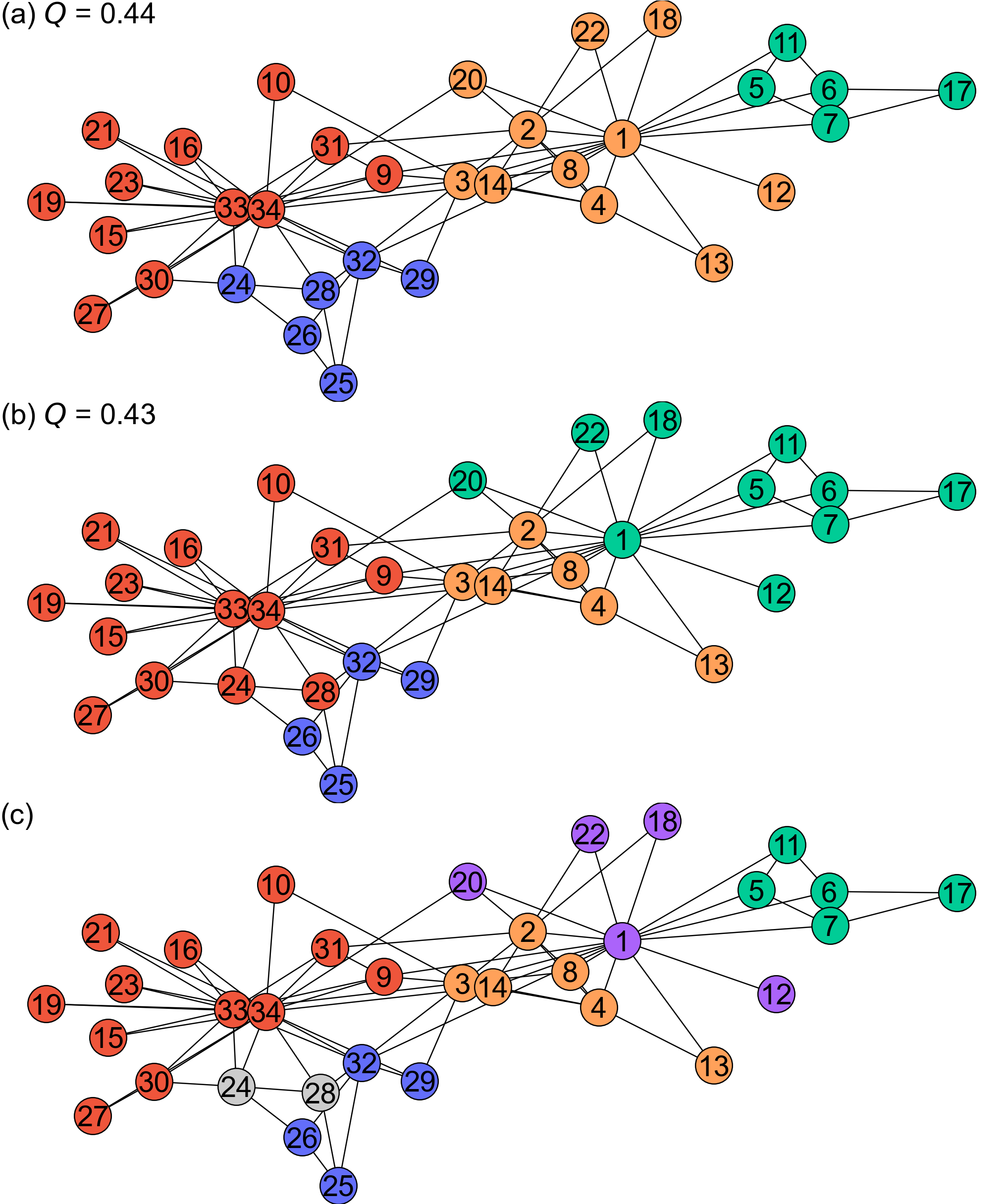}
    \caption{Delineating cores in Zachary's Karate Club weighted network.
    (a) and (b) Detected communities (shaded in different colours) using the Louvain method to maximize modularity with different random seeds.
    Despite having similar (and high) modularity scores $Q$, the community structure is degenerate.
    (c) Cores of the network partition ensemble generated using 1000 different random seeds.
    Cores are indicated by colour, with grey denoting nodes with unstable community assignments.
    It is plain to see that nodes in each core cluster together in (a) and (b), but with core $\{1, 12, 18, 20, 22\}$ having different community allegiance.
    The unstable nodes $\{24, 28\}$ change community assignments independently of each other and other nodes when considering the full 1000 random seeds.
    }
    \label{fig:zachary}
\end{figure}

To illustrate this problem, let us consider Zachary's Karate Club \cite{Zachary1977}, a social network ubiquitous in the study of community structure \cite{GirvanNewman2002,NewmanGirvan2004,Gfeller2005,Agarwal2008,Karrer2008,Cheng2010,Peel2017}.
Community detection in this network aims to partition the 34 members of a karate club based on observed social interactions, optionally weighted by the strength of ties between individuals.
If we attempt then to maximize modularity \cite{NewmanGirvan2004, Newman2006}, a prototypical measure of the quality of a division of a network into communities, using a common community detection algorithm such as the Louvain method \cite{Blondel2008}, we receive community descriptions of similar quality sensitive to random seed (\cref{fig:zachary}a, b).
If one accepts modularity as a suitable gauge of community structure quality, it is tempting to proceed with analyses on the partition with maximal modularity, assuming it is practical to find \cite{LancichinettiFortunato2011}.
It is unclear, though, what scientific value these analyses hold if the optimal partition's score is roughly equal to an exponential number of degenerate partitions \cite{Agarwal2008,Good2010}.

If a network exhibiting degenerate community structure makes the interpretation of a single partition misleading \cite{Massen2006, Good2010, Reijnders2021}, and consensus solutions can blur relevant information captured by different partitions \cite{Calatayud2019}, how can results be interpreted in a scientific context?
Ideally, the consistent features of the high-scoring community structures should be considered to gain insight into the structure of the network \cite{Riolo2020, Lee2021}.
We present here a recursive significance clustering scheme to reveal \emph{cores}: subsets of nodes that are jointly assigned to communities with significance (e.g., \cref{fig:zachary}c).
Cores are the features of network community structure that are well-supported by the data.

We first introduce the recursive significance clustering technique (\cref{sec:rsc}) and accompanying metrics on network cores (\cref{sec:metrics}).
We then demonstrate the analytical capabilities of this approach by example on a real network (\cref{sec:application}) and validate it by studying the stability of community structure in synthetic networks (\cref{sec:synthetic}).
All materials and methods used are described in full detail in \cref{sec:methods}.

\section{\label{sec:rsc} Recursive significance clustering}
For the results of community detection to be of value, the extent to which detected communities are statistically significant as opposed to coincidental must be considered \cite{Guimera2004,Karrer2008}.
After all, it is possible that a network has very weak or no community structure, meaning that detected communities are fragile and merely an artifact of the detection method used \cite{Good2010, Calatayud2019}.
A distinct optimal solution should be robust to small changes to the network's configuration owing to the gap in quality to competing optima.
In turn, the matter of community structure significance is reformulated as a question of how robust that structure is to network perturbations \cite{Karrer2008}.

To identify the consistent structure in an ensemble of partitions, \emph{significance clustering} procedures seek to distinguish between stable and unstable node assignments to communities \cite{RosvallBergstrom2010, Calatayud2019}.
Stable or significant nodes of a community are flagged based on the proportion of ensemble partitions in which they are jointly assigned to a community.
This approach discovers only the largest significant subset of nodes belonging to a community, but there may exist other mutually exclusive subsets of nodes in that community that are jointly assigned across the solution ensemble, but not with the largest or other subsets.
This limitation betrays another: a (single) reference network partition is used to identify these significant subsets, necessarily dictating the quantity and delineation of significant subsets found.
As a single partition neglects relevant information available in the whole solution landscape, this reliance on a reference partition as the mould for significant subsets is undesirable.

We define the robust cores of a network as the subsets of nodes that are jointly assigned to communities, capturing the consistent features of the landscape of community structures.
For significance level $\alpha$, we require that all nodes of a core are jointly assigned in a $(1-\alpha)$-fraction of partitions to relax the requirement.
The arrangement of cores into communities may vary among solutions, but that of nodes into cores is robust; in this light, cores are nearly indivisible constituents of community structure \cite{Riolo2020}.
Then, delineating cores is akin to assessing the community structure variability, as they are the invariant units that compose dissimilar partitions.
Nodes that do not regularly cluster with any other nodes, and nodes belonging to cores too small relative to the size of the network to contribute meaningful insight into system function, are regarded as unstable or noisy assignments.

The recursive significance clustering scheme we propose to identify network cores is outlined as follows, noting that the ways in which steps of this procedure are carried out are flexible to allow for operations germane to the network under study to be used:
\begin{enumerate}
\item The empirical network has its components perturbed repeatedly to generate many replicate networks that together capture modeled variability of and/or uncertainty in the network's configuration.
\item Community detection is performed on each network realization, forming an ensemble of network partitions.
\item An optimization technique is used to find the largest subset of nodes that are all assigned to the same community in a $(1-\alpha)$-fraction of partitions; these nodes are designated as a core, removed from the node pool, and then the next largest core is searched for, and so on until all cores of meaningful size are extracted.
\end{enumerate}
This recursive approach has the advantage of operating without a reference partition; instead, all detected community structures are intrinsically used to discover the well-supported features of network structure.
\Cref{sec:methods} provides a detailed account of how we implement each step.

\section{\label{sec:metrics} Viewing community structure variability}
We have established cores as the elemental units of nodes that (i) are arranged like building blocks to form communities and (ii) are nearly indivisible.
Identifying exactly how these cores coalesce and how stable they are as a unit provides a direct view of structural patterns in a degenerate community structure landscape and, thus, illuminates the robust features of network structure.

For this, we define the \emph{coalescence frequency} of a union of cores to be the proportion of partitions in which the nodes of those cores are all assigned to the same community and without other cores.
Notably, the coalescence frequency of one core is equal to the proportion of partitions where that core does not merge with other cores, instead consistently forming a community by itself.

In addition, the \emph{stability} of a core is the proportion of partitions in which the nodes of the core are jointly assigned to a community.
By definition, it is guaranteed to be within the interval $[1-\alpha, 1]$ for significance level $\alpha$.
The sum of every coalescence frequency of core unions including some core will equal that core's stability.

\section{\label{sec:application} Application}
To illustrate this approach, we analyze the cores of a real network.
We begin by introducing the network and the recursive significance clustering setup (\cref{sec:application-net}).
We then examine how the network cores reveal robust structure in the network (\cref{sec:application-cores}) and how this structure is cohesive over its temporal evolution (\cref{sec:application-temporal}), and briefly examine the nature of nodes with unstable community assignments (\cref{sec:application-noise}).

\subsection{\label{sec:application-net} Alaska--Alaska flight network}
Flight networks benefit from community detection to uncover the structure of flight routes \cite{Guimera2005,Cheung2020}.
The network of flights in Alaska (AK), United States, serves as an interesting case study due to the key role that flight plays in the region's transportation system \cite{Conway2005}.

We consider three temporal snapshots of the AK--AK flight network: calendar years 2017, 2020, and 2023.
In the 2017 (resp. 2020; 2023) network, there are 3,578 (3,285; 3,616) directed links weighted by the number of aggregate flights in that year from one of the 334 (328; 330) airports to another, for 401,153 (275,890; 364,285) total annual flights.
(See \cref{sec:materials} for information on data compilation). 
These years were selected to study flight traffic before, during, and after the largest impacts of the COVID-19 pandemic, a major disruption in aviation activity \cite{Sun2022}, which affords a prime opportunity to analyze the temporal behaviour of robust network cores.

We assume flights are independent events and resample each link's weight from a Poisson distribution with rate parameter equal to that link's weight in the empirical network to form 1000 replicate networks \cite{RosvallBergstrom2010}.
We then find the robust cores of the network's community structure, detected using Infomap \cite{Edler2017}, at a significance level of 0.05, accepting only cores of at least five nodes.
This is completed for each temporal snapshot.
(See \cref{sec:methods} for a detailed description of this procedure.)

\subsection{\label{sec:application-cores} Reckoning with dissimilar structure}
\begin{figure}
    \centering
    \includegraphics{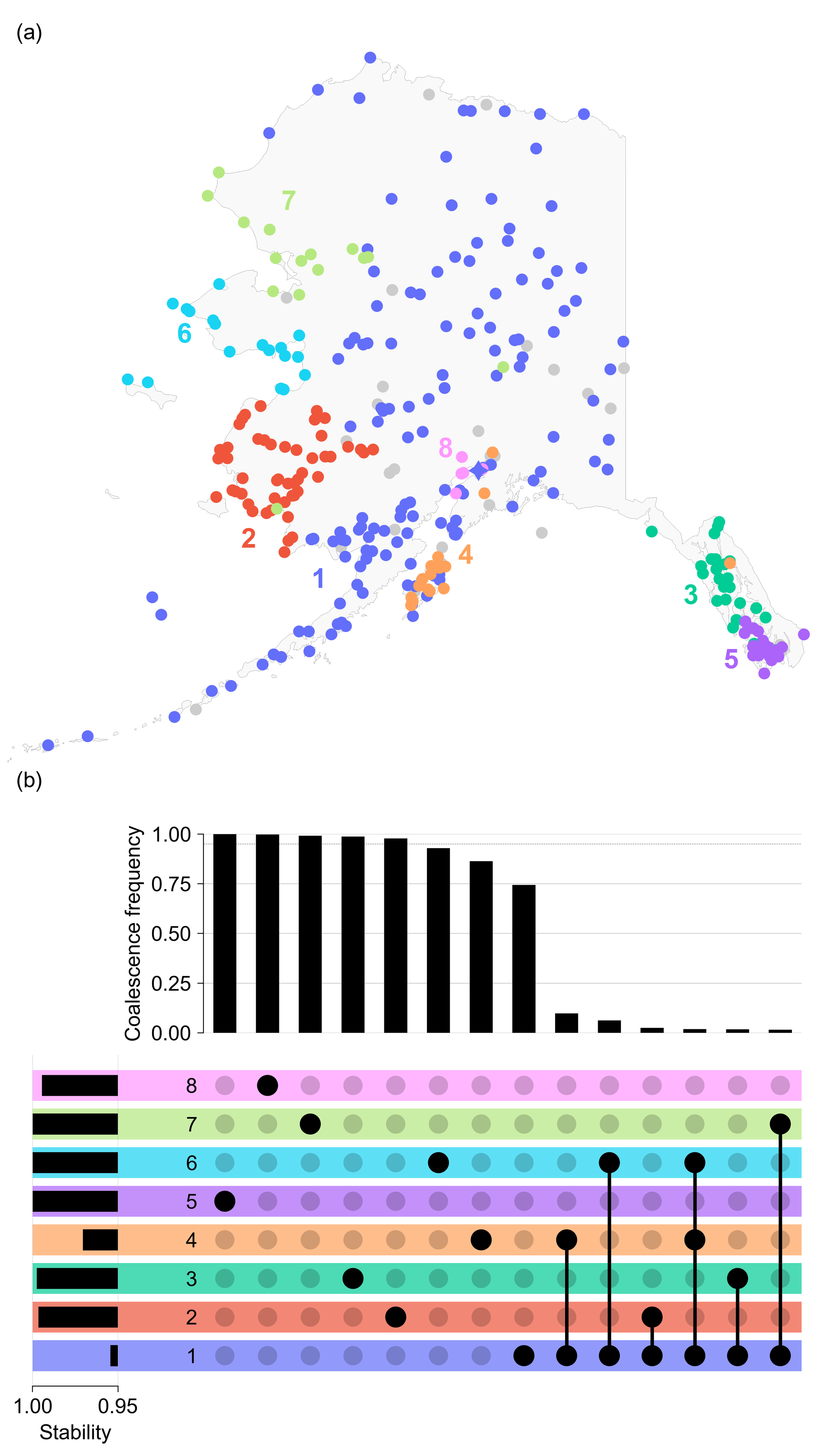}
    \caption{Cores of the 2017 temporal snapshot of the AK--AK flight network.
    (a) Spatially embedded cores.
    Nodes with unstable community assignments are shaded grey.
    (b) Core coalescence frequency and stability.
    }
    \label{fig:alaska2017}
\end{figure}

Spatially embedded cores of the 2017 AK--AK flight network snapshot with their coalescence frequencies and stability are shown in \cref{fig:alaska2017} (find that of the 2020 and 2023 snapshots in \cref{app:snapshots}). 
For visualization, we modify the UpSet plot \cite{Lex2014} to show the coalescence frequency for any combination of cores in tandem with the stability of each core.
To manage possible combinatorial explosion, only core combinations with a coalescence frequency of at least 0.05 are displayed.

From \cref{fig:alaska2017}, we can appreciate that the AK--AK flight network has very consistent community structure, with each core forming its own community much more frequently than not.
When cores do unite as communities, it always involves the large (by geographic span and node count) core 1.
Speaking to the system, core 1 includes the Ted Stevens Anchorage International Airport (IATA: ANC; marked with a star), which brokers a tremendous volume of AK flight traffic, making it a highly influential node in the network as measured by virtually any centrality index.
This is responsible for the vast geographic distribution of core 1 compared to other cores: the connectivity of ANC recruits airports to core 1 that otherwise have no strong local connection structure.
The coalescence frequencies involving core 1 show that it sometimes merges with cores 2, 3, 4, 6, and 7.
We imagine this as it irregularly absorbing the smaller cores in close proximity to it, indicating some presence of rather coarse community structure engulfing the west coast and much of the inland in the ensemble.

\subsection{\label{sec:application-temporal} Temporal cohesion of cores}
\begin{figure}
    \centering
    \includegraphics{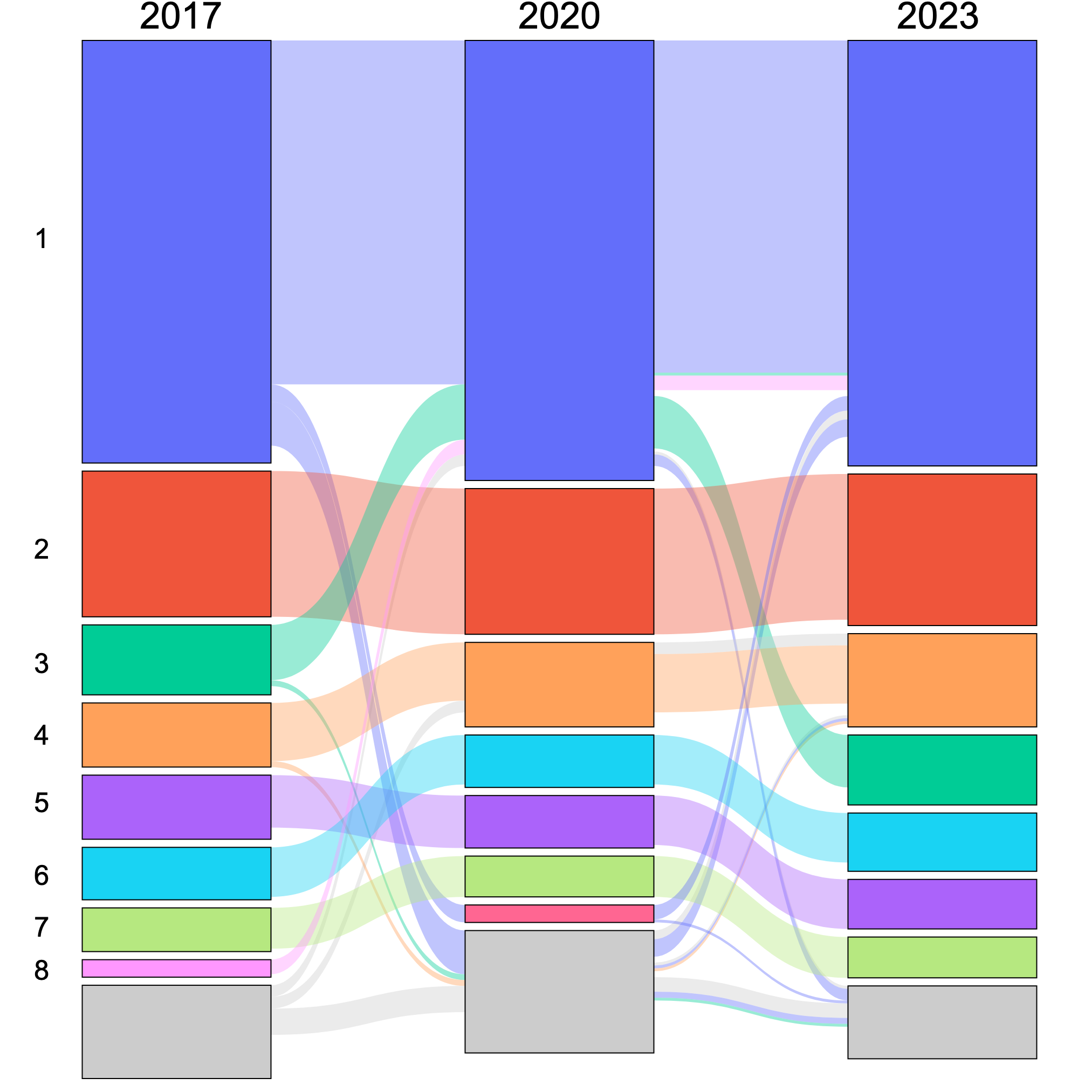}
    \caption{Core assignment changes of nodes through the 2017, 2020, and 2023 temporal snapshots of the AK--AK flight network.
    Nodes with unstable community assignments are placed in the grey-shaded block.
    }
    \label{fig:alluvial}
\end{figure}

Now we examine the cores of successive network snapshots to assess structural changes in the network over time.
An alluvial diagram \cite{RosvallBergstrom2010} shows this in \cref{fig:alluvial}.
Streamlines show the core assignments of nodes changing between snapshots, coloured by their initial core assignment to track them in time.
Since the node set of the network is not static in time, the flow of a node assignment is not shown if the node is not present in a contiguous temporal snapshot.

It is immediately apparent that each core is largely cohesive in time, i.e., cores continue to be stable through successive temporal snapshots.
Following the streamlines, we see that some nodes change their stability with flight volume.
Otherwise, the only core dynamics are the absorbing of cores 3 and 8 to core 1 in 2020 --- with core 3 splitting shortly thereafter in 2023 --- and core 1 ejecting a new core in 2020, which then rejoins core 1 in 2023.

An anomalous change in flight volume between an individual airport and ANC can obfuscate or reveal small-scale substructure due to the influence of ANC in the network.
An increase in flight volume from 2017 occurred between ANC and Beluga Airport (BVU) of core 9 to catalyze the merge with core 1 in 2020, and a decrease between ANC and Homer Airport (HOM) for the emergence of the smaller core split from core 1.
The core size threshold could be adjusted \textit{post hoc} to exclude these small cores if their inclusion is felt to be more distracting than informative.

\subsection{\label{sec:application-noise} Speculating on the nature of unstable nodes}
\begin{figure*}
    \centering
    \includegraphics{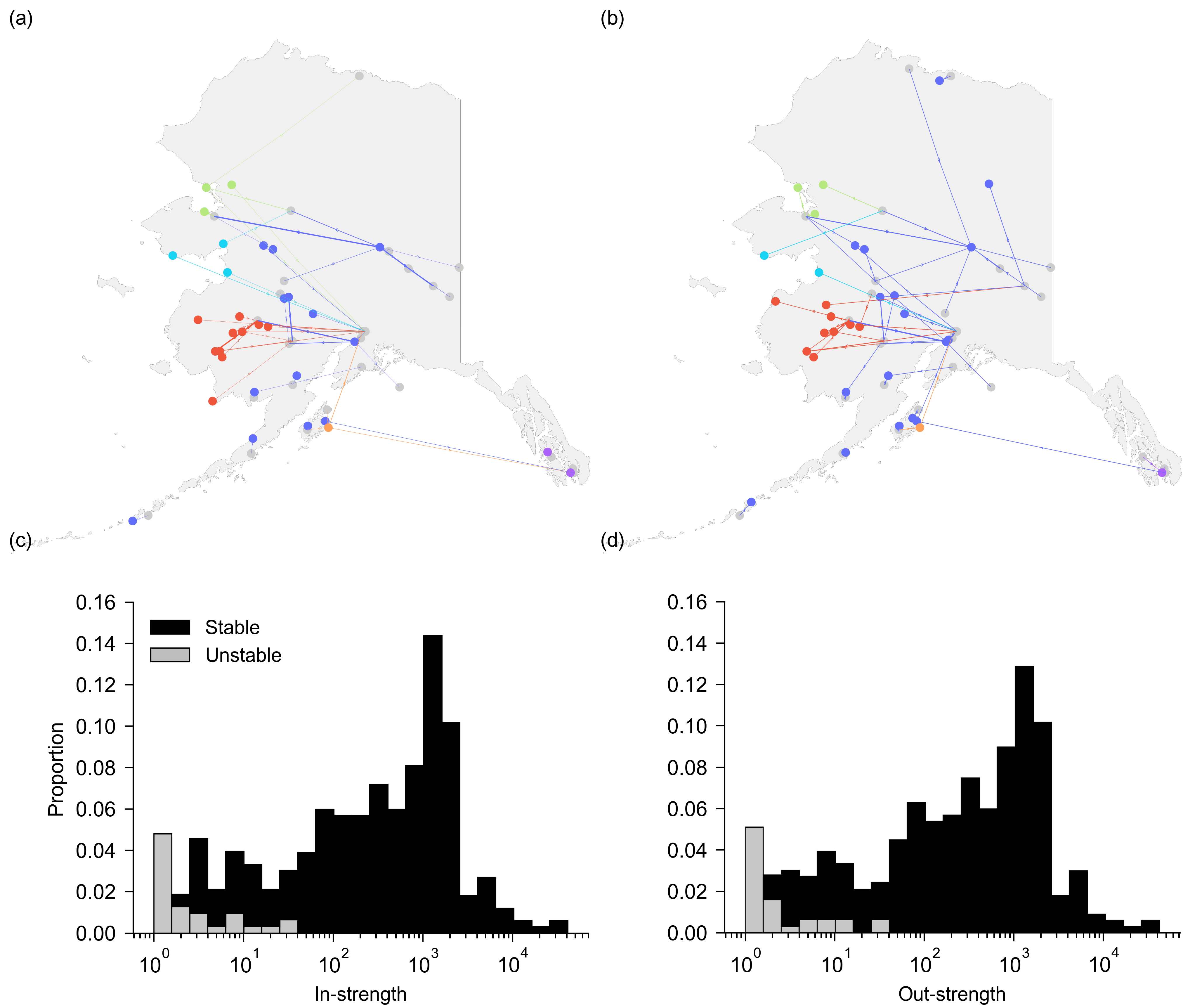}
    \caption{
    Wiring architectures and strength distributions of nodes with unstable community assignments for the 2017 AK--AK flight network.
    (a) and (b) In- and out-links, respectively, depicting the interactions of spatially embedded unstable nodes with all other nodes. 
    Link widths are proportional to link weight. 
    (c) and (d) Distribution of in- and out-strengths, respectively, of unstable nodes compared to that of stable nodes.
    }\label{fig:noise}
\end{figure*}

We now turn our focus away from cores and toward the residual nodes with unstable assignments found in the recursive significance clustering process.
While we leave a comprehensive investigation into the nature of nodes with unstable community assignments to future studies, we conjecture a classification of their role in network connectivity based on observations in the study of the AK--AK flight network.

On average, unstable nodes have substantially smaller in- and out-strengths than the stable nodes (\cref{fig:noise}c, d), corresponding to aerodromes with fewer arrivals and departures.
Looking at the origins and destinations of these comparatively few flights (\cref{fig:noise}a, b), we can loosely place each unstable node into one of two functional categories \cite{Kim2019}: (i) a leaf node having low connectivity with a single stable node, or (ii) a branch node with connections to multiple stable nodes belonging to different cores.

The unstable-leaf nodes are unstable in the sense that they are weakly connected to one core and, as such, are frequently assigned to their own singleton communities in ensemble partitions.
It could reasonably be argued that assigning these leaf nodes to their adjacent cores (i.e., the core to which the node they are linked to belongs) is a more apt depiction of mesoscale system function.
However, this is not to diminish the value of marking them as unstable, as one should be cautious of relying on these individual nodes to be jointly assigned with other nodes of their adjacent cores because of this low connectivity.

On the other hand, unstable-branch nodes have community assignments that are somewhat independent from that of their adjacent cores, leading to their interpretation as the uncertain (fuzzy) borders between communities.
We recognize that in many real networks, a hierarchical community structure is a preferable portrayal of network organization compared to the simplified non-overlapping representation \cite{Palla2005,Leskovec2009,YangLeskovec2014,Jeub2015}.
Unstable-branch nodes may be better described as nodes having joint membership to multiple (overlapping) communities in hierarchical structure.

\section{\label{sec:synthetic} Community structure stability}
While the AK--AK flight network demonstrated very strong community structure, it is important to characterize the cores found --- and not found --- in networks with weak or no community structure.
The strength of the community structure in real networks can range from strong (few intercommunity links) to weak (many intercommunity links) \cite{Ghalmane2019}.
We make use of the Lancichinetti--Fortunato--Radicchi (LFR) benchmark network \cite{LFR2008}, which is generated in the likeness of real networks, exhibiting many of the same common properties, including planted community structure.
We pay particular attention to the \emph{mixing parameter} $\mu$.
Since each node shares a $(1-\mu)$-fraction of its links with its own community and a $\mu$-fraction with the rest of the network, $\mu \in [0,1]$ can be considered a slider with which to tune the strength of the community structure.

Ideally, we should find a large proportion of nodes in the network to be unstable when $\mu$ is large, and most nodes stable at small $\mu$.
This would reflect the lack of clear community structure when there is greater mixing between communities.
Beyond the critical point $\mu = 0.5$, communities are no longer defined in the strong sense, with each node sharing fewer links with its community than the rest of the network \cite{Radicchi2004}.
In general, the success in recovering planted communities through modularity maximization has been shown to decline steeply approaching this threshold and beyond \cite{LFR2008,Yang2016}.
We detect communities in undirected and unweighted LFR benchmark networks by maximizing modularity with the Louvain method \cite{Blondel2008}.
Since we are concentrated on the positioning of links, we perturb the network by randomly redirecting a small percentage of its links.
Recursive significance clustering is carried out as in the AK--AK flight network. (See \cref{sec:methods}.)

\begin{figure}
    \centering
    \includegraphics{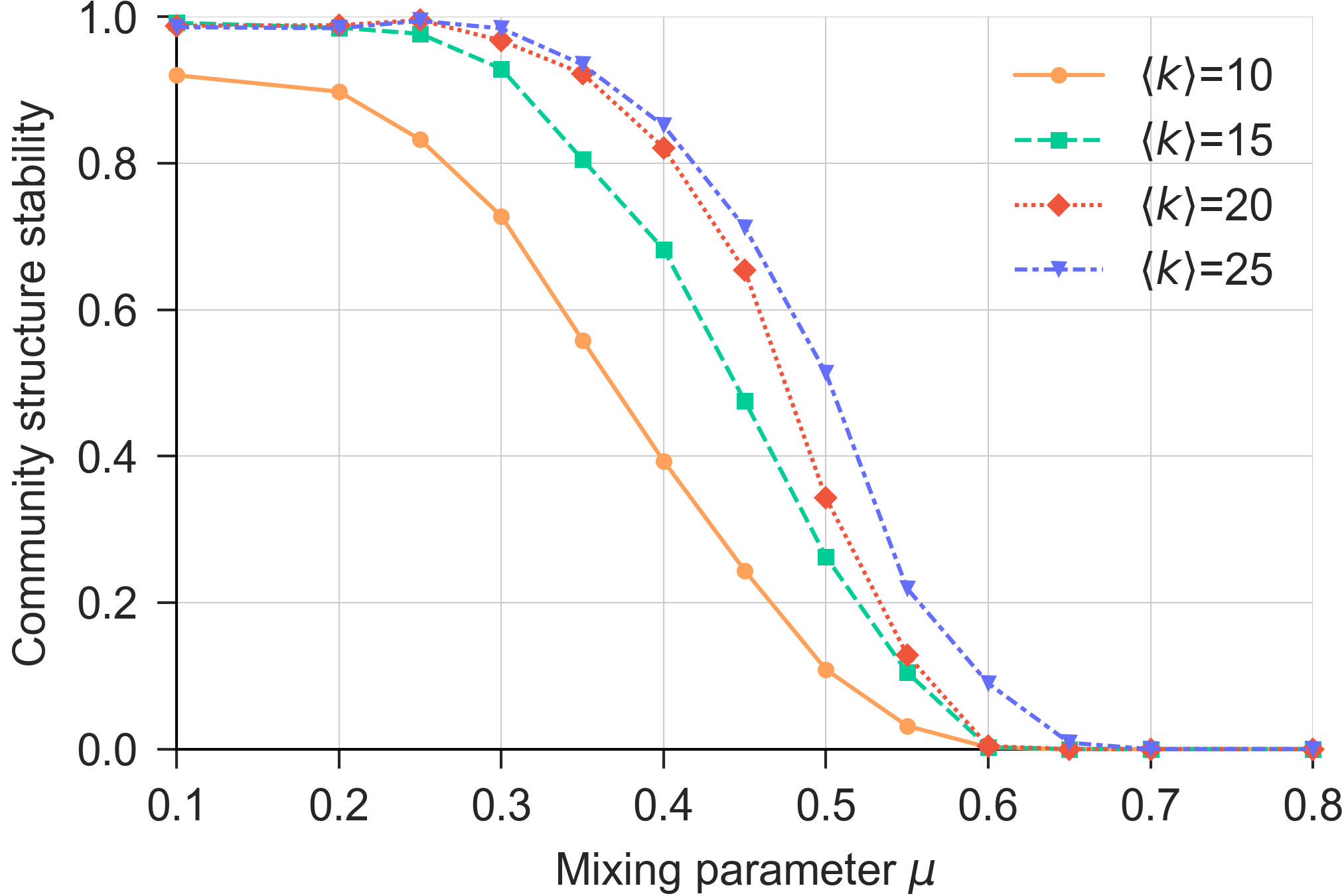}
    \caption{
    Proportion of nodes with stable community assignments in LFR benchmark networks with  mixing parameter $\mu$ and average degree $\langle k \rangle$.
    }
    \label{fig:stability}
\end{figure}

To summarize the content of \cref{fig:stability}, LFR benchmark networks with higher link density $\langle k \rangle$ are more stable at the same level of mixing $\mu$.
This implies that the community structure in a network with sparse link density is more sensitive to link rearrangement, even when the number of links redirected is proportional to this density.

At low mixing, at most one core is associated with each planted community.
When community structure is defined only in the weak sense beyond $\mu = 0.5$, the structure is highly unstable.
Between the low and high mixing regimes is a phase transition in which increasingly more planted communities are unstable.
That is, entire communities of nodes become the so-called unstable-branch nodes in the network.
This trait also partially accounts for a baseline instability at low mixing, where the randomly generated wiring of one or more (comparatively small) communities is not robust.
This is in addition to the presence of unstable-leaf nodes that are easily severed from their community at all levels of mixing, but more prominent in networks with sparser link density.

We point out the stability plateau that emerges around $\mu = 0.25$ for the two networks with higher link density $\langle k \rangle = 20, 25$.
This increase in stability with $\mu$ from  baseline instability is an exception to the monotonic non-increasing behaviour of the curves.
It could suggest that, for denser networks, some degree of mixing actually fortifies the network structure against rewiring.
A study of this phenomenon is left to future work.

We see then from this experiment that the existence of cores in a network is symptomatic of robust community structure.
This is in a local sense: significant community structure in one part of a network is identified independently of the significance of structure elsewhere in the network and without a reference partition prescribing the locality.

\section{\label{sec:conclusions} Conclusions}
Here we have introduced a recursive significance clustering scheme to identify the robust features the variability of community structure in complex networks.
We extracted the cores of networks as the subsets of nodes that have stable joint community assignments under network perturbation.
Cores serve as atomic building blocks that are arranged in different configurations to form a  community structure landscape but are nearly indivisible themselves.
Based on these properties of cores, we defined the core coalescence frequency and stability metrics as the frequency that cores unite into communities and nodes into cores, respectively.
This method also inherently distinguishes the nodes with noisy community assignments, allowing them to be filtered to facilitate meaningful comparisons of network partitions.

Studies using community detection to guide a scientific pursuit will benefit from a cognizance of cores to ensure analyses are not conducted on only one network partition of only one network sample, and instead embrace uncertainty or noise in the network's configuration to demarcate the robust features of its structure.
We note that while our recursive significance clustering technique has been presented here in the context of complex network community structure, its premise is perfectly suitable for an examination of any clustered data.
It would benefit from future development extending its use to applications of hierarchical clustering.

We look forward to a full comparison of network cores from recursive significance clustering and consensus clustering solutions in future work.
We foresee the use of cores as building blocks to form a consensus solution of robust community structure according to the needs of one's application.

\section{\label{sec:methods} Materials and methods}
\subsection{Network}
\label{sec:materials}
We mathematically formulate each network as $G = \left(V, E, W\right)$, where $V$ is the network's set of nodes, $E \subseteq V \times V$ is its set of (directed) links, and $W \colon E \to \mathbb{R}^+$ is a weighting function that assigns positive and real weights to links.
When the network is unweighted, the weight of each edge is equal to one.
When it is undirected, a link $e_{u,v} \in E$ between nodes $u,v \in V$ is equivalent to the link $e_{v,u}$.
To simplify notation, some link that is not a component of the network $e_{u,v} \not\in E$ is assigned a weight of zero $W(e_{u,v}) = 0$.

\paragraph*{AK--AK flight network}
The described AK--AK flight network was constructed from the 2017, 2020, and 2023 T-100 Domestic Segment Data from the US Bureau of Transportation Statistics. 
Entries are flight segments, defined as a pair of domestic US airports served by a single stage of at least one flight during the reporting year.
For a given year, we consider only the airports in AK with at least one flight to or from another airport in AK.
We use IATA, FAA, and ICAO identifiers to obtain airport GPS coordinates.
Each temporal snapshot of the AK--AK flight network is a separate network with nodes representing aerodromes and directed edges flight segments weighted by annual flight count.

\paragraph*{Synthetic network}
We generate LFR benchmark networks \cite{LFR2008} with variable mixing parameter $\mu$ and average degree $\langle k \rangle$.
With 1000 nodes, we set the maximum degree and community size equal to 100 to practically synthesize the network \cite{Yang2016, Tian2023}.
We use degree and community size distribution exponents of -3 and -2, respectively, which are typical values found in real networks \cite{LFR2008}, including the AK--AK flight network.
We generate three networks with different random seeds per parameterization and report the mean of results.

\subsection{Network perturbation}
We aim now to somehow perturb the atomic components of $G$ repeatedly to generate a collection of $N = 1000$ replicate networks $\mathcal{G} = \{G^*_1,\ldots,G^*_N\}$.
For many complex networks, it makes little sense to discuss the network having a different set of nodes.
Zachary's Karate Club, for example, is an idiosyncratic network that would be unrecognizable without the instructor or with an additional officer; the same is true for our flight network.
So as to not undermine the individual characteristics of the nodes, we modify the relationships between them, that is, links \cite{RosvallBergstrom2010}.

\paragraph*{AK--AK flight network} Convenient for networks that preclude replicate observations, we make use of the statistical bootstrap \cite{Efron1979} to produce replicate networks by parameterizing and resampling link weights \cite{RosvallBergstrom2010}.
Treating flights as independent events, we resample each link's weight from a Poisson distribution with a rate parameter equal to that link's weight in the empirical network.
In this manner, each replicate network $G^*_i = \left(V, E, W^*_i\right)$ varies by weighting function, where the weight of each link $e \in E$ is i.i.d. such that $W^*_i(e) \sim \mathrm{Pois}(W(e))$.

\paragraph*{Synthetic network}
As we examine the stability of community structure as a function of mixing $\mu$, which is concerned with the position of links between and within communities, we perturb each LFR benchmark network by rewiring a small percentage of its links.
For $\Sigma$ undirected and unweighted links, we randomly draw $0.05\Sigma$ links to redirect.
One endpoint of each drawn edge is replaced with another node at random; if such an edge already exists in the network it is simply removed since the network is unweighted.

We stress that a plethora of other perturbation/resampling methods are acceptable.
For example, introducing additive noise over link weights has the advantage of being directly interpreted as uncertainty in the empirical network configuration in many contexts \cite{Gfeller2005}.
Links may be repositioned in alternative fashions \cite{Karrer2008}, added \cite{Tian2023}, or randomly sampled \cite{Costenbader2003}, and other suitable methods may be used to model variability in the networked system.
Indeed, if many replicate observations of the network were available, these could be used in lieu of generating artificial networks.

In the introductory example of Zachary's Karate Club (\cref{fig:zachary}), instead of perturbing the network to generate a partition ensemble, we effectively perturbed the stochastic community detection algorithm operating on a static network through the use of different random seeds \cite{Kwak2009}.
Although beyond the scope of this work, we acknowledge the potential utility of this method in identifying common structure among results from different community detection strategies.

\subsection{Community detection}
We now require a community description $M^*_i$ that is a partition of the node set $V$ for each network realization $G^*_i \in \mathcal{G}$ to generate a set of partitions $\mathcal{M} = \{M^*_1,\ldots,M^*_N\}$.
We consider non-overlapping as opposed to hierarchical community structure such that each node is assigned to one and only one community in each community description for ease of interpretation of results.
The community detection algorithm that best fits the network under study should be used \cite{Fortunato2010}.

\paragraph*{AK--AK flight network}
Leveraging that our flight network is both directed and weighted, and can be  analogized as a model of air traffic flow, we use the stochastic search algorithm Infomap \cite{Edler2017} making use of the map equation framework \cite{RosvallBergstrom2008} to detect community structure.
The map equation quantifies the quality of a network partition by its ability to compress information flow on the network as a means of identifying regularities in network structure with respect to flow \cite{Rosvall2009}.
Communities under the map equation regime then correspond to structures of persistent flow, implying strong internal mixing and coherence \cite{Ser-Giacomi2015}.
To tune detected communities to a desirable scale for analysis, we set a variable Markov time \cite{Schaub2012a,Schaub2012b,Edler2022} of 3.

\paragraph*{Synthetic network}
The modularity of a partition $M$ of the undirected network $G$ is 
\begin{equation*}
    Q(G, M) = \frac{1}{2\Sigma}\sum_{u,v \in V}\left(W(e_{u,v})-\frac{k_uk_v}{2\Sigma}\right)\delta(m_u,m_v),
\end{equation*}
where $\Sigma$ is the sum of link weights in the network, $k_u$ is the strength of node $u$, and $\delta$ is the Kronecker delta function such that $\delta(m_u, m_v)$ evaluates to unity if nodes $u$ and $v$ are jointly assigned to a community in $M$ and zero otherwise \cite{CNM2004}.
The Louvain method of community detection finds a partition $M$ that maximizes $Q$ through a greedy stochastic search \cite{Blondel2008}.

\subsection{Core search}
We wish to find the cores of the network $G$ with significance level $\alpha = 0.05$, i.e., subsets of nodes that are jointly assigned to the same community over a $(1-\alpha)$-fraction of community descriptions in the solution ensemble $\mathcal{M}$.
This is done in a recursive manner, such that the largest core of the network is identified first, then the next largest, and so on until all desired cores are delineated.
To define what is meant by the "largest" core, we must specify a measure of size of a subset of nodes $\nu: 2^V \to \mathbb{R}^+$, which we take simply to be the cardinality of the set.
This choice, in essence, treats each node as having equal importance in the delineation of cores.

Following the seminal work on significance clustering of \cite{RosvallBergstrom2010}, the score of an arbitrary candidate core $C \subseteq V$ is denoted $s(C)$ and is the size of the core $\nu(C)$ minus a penalty to ensure its nodes cluster together over a $(1-\alpha)$-fraction of ensemble solutions.
To calculate the penalty, for every community description $M'_i \in \mathcal{M}$, we find the community $m \in M'_i$ that contains the most nodes of the candidate core $C$.
The mismatch nodes $C \setminus m$ that are assigned to $C$ but not to $m$ break the constraint that the core must cluster together.
The size of this mismatch node set $\nu(C \setminus m)$ in every community description is summed, excepting the $\alpha$-fraction producing the largest penalties to relax the requirement.
This quantity is the penalty associated with $C$.

To approximate the largest core, we employ simulated annealing \cite{Kirkpatrick1983}.
To begin, on sweep $t = 0$ with temperature $T_0 = 1$, a candidate core $C \subseteq V$ is randomly initialized.
A node $v \in V$ is drawn at random, and has its membership to the candidate core $C$ flipped to create a trial core $C'$.
Per the Metropolis--Hastings algorithm \cite{Metropolis1953,Hastings1970}, this trial core is accepted as the candidate core $C \leftarrow C'$ with probability equal to $\min\left\{1,\, \exp((s(C') - s(C))/T_t)\right\}$.

The generation of candidate cores is repeated with replacement of drawn nodes as many times as there are nodes to draw from to complete one full sweep.
After a full sweep, the temperature is cooled on an exponential schedule $T_t = T_0r^t$, where $r \in \left(0, 1\right)$ is a cooling rate parameter we set \textit{ad hoc} to $r=0.99$.
Cooling continues until no trial core is accepted on a sweep, at which point the candidate core $C$ is taken as an approximation to the largest core.

We repeat the search for the largest core recursively. 
After a core is found, we exclude all nodes belonging to it from the pool of eligible nodes to be drawn from in subsequent searches (i.e., $V \setminus C$ becomes the new node pool) and repeat the simulated annealing.
The search ends when a core is found that is not of an appreciable size relative to the network.
That core, and any nodes still not assigned to any core, are dubbed noise.
Before finishing the process, unassigned nodes are riffled through to see if they may be assigned to any core, checked in descending order of core size, to manually correct non-exhaustive searches.

The simulated annealing scheme used here is relatively primitive compared to the modern search algorithms used in community detection.
Its run-time performance can be improved to scale more efficiently with network size by incorporating techniques from the literature \cite{Fortunato2010}, most valuably heuristic candidate generation and restarts.

\appendix
\section{\label{app:snapshots} Temporal snapshots}
\begin{figure}
    \centering
    \includegraphics{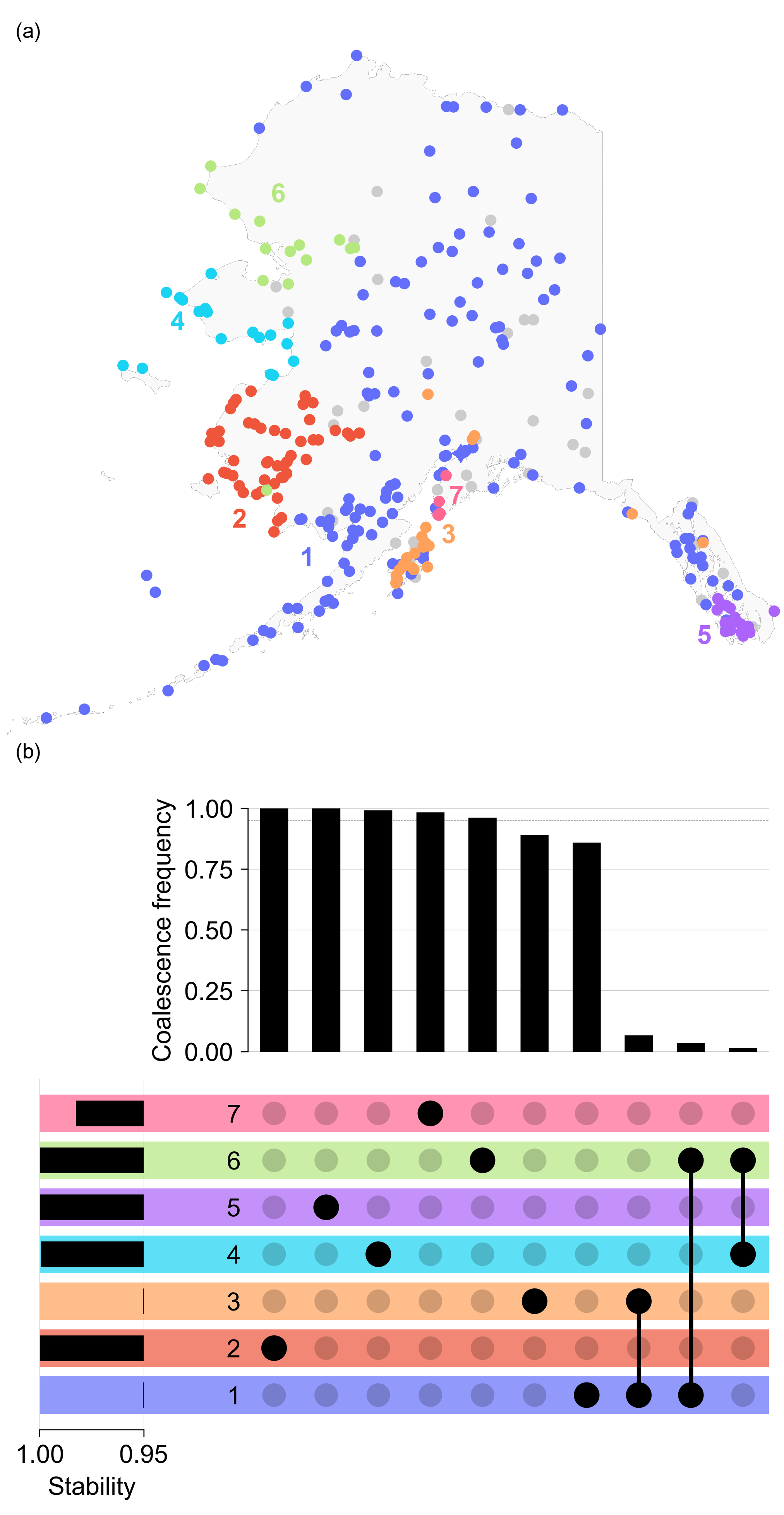}
    \caption{Cores of the 2020 temporal snapshot of the AK--AK flight network.
    (a) Spatially embedded cores.
    Nodes with unstable community assignments are shaded grey.
    (b) Core coalescence frequency and stability.
    }
    \label{fig:alaska2020}
\end{figure}

\begin{figure}
    \centering
    \includegraphics{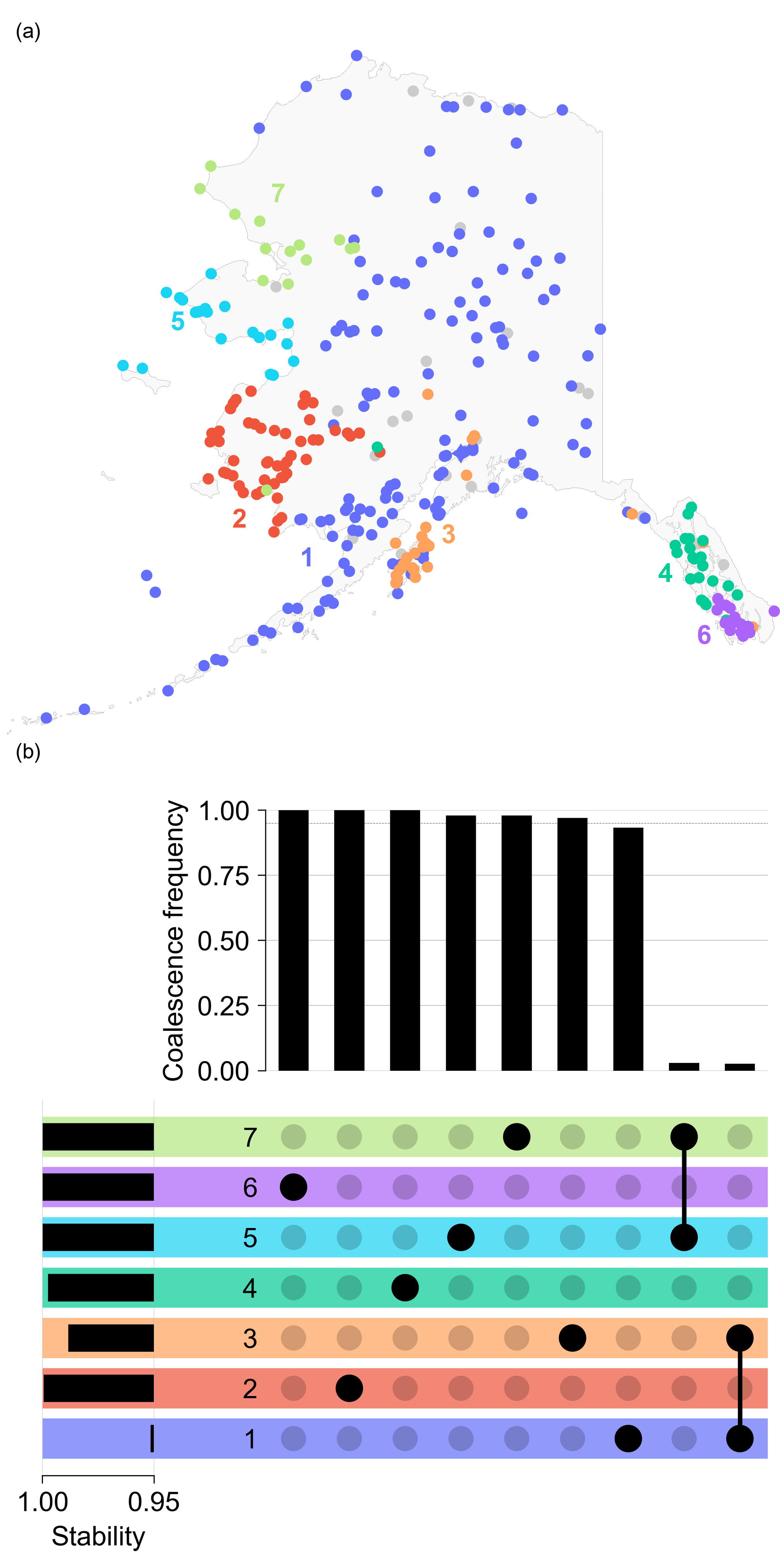}
    \caption{Cores of the 2023 temporal snapshot of the AK--AK flight network.
    (a) Spatially embedded cores.
    Nodes with unstable community assignments are shaded grey.
    (b) Core coalescence frequency and stability.
    }
    \label{fig:alaska2023}
\end{figure}

Cores of the 2020 and 2023 temporal snapshots of the AK--AK flight network are shown in \cref{fig:alaska2020} and \cref{fig:alaska2023}, respectively.

\bibliography{refs}

\begin{thebibliography}{62}%
\makeatletter
\providecommand \@ifxundefined [1]{%
 \@ifx{#1\undefined}
}%
\providecommand \@ifnum [1]{%
 \ifnum #1\expandafter \@firstoftwo
 \else \expandafter \@secondoftwo
 \fi
}%
\providecommand \@ifx [1]{%
 \ifx #1\expandafter \@firstoftwo
 \else \expandafter \@secondoftwo
 \fi
}%
\providecommand \natexlab [1]{#1}%
\providecommand \enquote  [1]{``#1''}%
\providecommand \bibnamefont  [1]{#1}%
\providecommand \bibfnamefont [1]{#1}%
\providecommand \citenamefont [1]{#1}%
\providecommand \href@noop [0]{\@secondoftwo}%
\providecommand \href [0]{\begingroup \@sanitize@url \@href}%
\providecommand \@href[1]{\@@startlink{#1}\@@href}%
\providecommand \@@href[1]{\endgroup#1\@@endlink}%
\providecommand \@sanitize@url [0]{\catcode `\\12\catcode `\$12\catcode `\&12\catcode `\#12\catcode `\^12\catcode `\_12\catcode `\%12\relax}%
\providecommand \@@startlink[1]{}%
\providecommand \@@endlink[0]{}%
\providecommand \url  [0]{\begingroup\@sanitize@url \@url }%
\providecommand \@url [1]{\endgroup\@href {#1}{\urlprefix }}%
\providecommand \urlprefix  [0]{URL }%
\providecommand \Eprint [0]{\href }%
\providecommand \doibase [0]{https://doi.org/}%
\providecommand \selectlanguage [0]{\@gobble}%
\providecommand \bibinfo  [0]{\@secondoftwo}%
\providecommand \bibfield  [0]{\@secondoftwo}%
\providecommand \translation [1]{[#1]}%
\providecommand \BibitemOpen [0]{}%
\providecommand \bibitemStop [0]{}%
\providecommand \bibitemNoStop [0]{.\EOS\space}%
\providecommand \EOS [0]{\spacefactor3000\relax}%
\providecommand \BibitemShut  [1]{\csname bibitem#1\endcsname}%
\let\auto@bib@innerbib\@empty
\bibitem [{\citenamefont {Strogatz}(2001)}]{Strogatz2001}%
  \BibitemOpen
  \bibfield  {author} {\bibinfo {author} {\bibfnamefont {S.~H.}\ \bibnamefont {Strogatz}},\ }\bibfield  {title} {\bibinfo {title} {Exploring complex networks},\ }\href {https://doi.org/10.1038/35065725} {\bibfield  {journal} {\bibinfo  {journal} {Nature}\ }\textbf {\bibinfo {volume} {410}},\ \bibinfo {pages} {268} (\bibinfo {year} {2001})}\BibitemShut {NoStop}%
\bibitem [{\citenamefont {Albert}\ and\ \citenamefont {Barabási}(2002)}]{AlbertBarabasi2002}%
  \BibitemOpen
  \bibfield  {author} {\bibinfo {author} {\bibfnamefont {R.}~\bibnamefont {Albert}}\ and\ \bibinfo {author} {\bibfnamefont {A.-L.}\ \bibnamefont {Barabási}},\ }\bibfield  {title} {\bibinfo {title} {Statistical mechanics of complex networks},\ }\href {https://doi.org/10.1103/revmodphys.74.47} {\bibfield  {journal} {\bibinfo  {journal} {Reviews of Modern Physics}\ }\textbf {\bibinfo {volume} {74}},\ \bibinfo {pages} {47–97} (\bibinfo {year} {2002})}\BibitemShut {NoStop}%
\bibitem [{\citenamefont {Newman}(2003)}]{Newman2003}%
  \BibitemOpen
  \bibfield  {author} {\bibinfo {author} {\bibfnamefont {M.~E.~J.}\ \bibnamefont {Newman}},\ }\bibfield  {title} {\bibinfo {title} {The structure and function of complex networks},\ }\href {https://doi.org/10.1137/S003614450342480} {\bibfield  {journal} {\bibinfo  {journal} {SIAM Review}\ }\textbf {\bibinfo {volume} {45}},\ \bibinfo {pages} {167} (\bibinfo {year} {2003})}\BibitemShut {NoStop}%
\bibitem [{\citenamefont {Barab{\'a}si}\ and\ \citenamefont {P{\'o}sfai}(2016)}]{Barabasi2016}%
  \BibitemOpen
  \bibfield  {author} {\bibinfo {author} {\bibfnamefont {A.-L.}\ \bibnamefont {Barab{\'a}si}}\ and\ \bibinfo {author} {\bibfnamefont {M.}~\bibnamefont {P{\'o}sfai}},\ }\href@noop {} {\emph {\bibinfo {title} {Network Science}}}\ (\bibinfo  {publisher} {Cambridge University Press},\ \bibinfo {year} {2016})\BibitemShut {NoStop}%
\bibitem [{\citenamefont {Girvan}\ and\ \citenamefont {Newman}(2002)}]{GirvanNewman2002}%
  \BibitemOpen
  \bibfield  {author} {\bibinfo {author} {\bibfnamefont {M.}~\bibnamefont {Girvan}}\ and\ \bibinfo {author} {\bibfnamefont {M.~E.~J.}\ \bibnamefont {Newman}},\ }\bibfield  {title} {\bibinfo {title} {Community structure in social and biological networks},\ }\href {https://doi.org/10.1073/pnas.122653799} {\bibfield  {journal} {\bibinfo  {journal} {Proceedings of the National Academy of Sciences}\ }\textbf {\bibinfo {volume} {99}},\ \bibinfo {pages} {7821} (\bibinfo {year} {2002})}\BibitemShut {NoStop}%
\bibitem [{\citenamefont {Fortunato}(2010)}]{Fortunato2010}%
  \BibitemOpen
  \bibfield  {author} {\bibinfo {author} {\bibfnamefont {S.}~\bibnamefont {Fortunato}},\ }\bibfield  {title} {\bibinfo {title} {Community detection in graphs},\ }\href {https://doi.org/10.1016/j.physrep.2009.11.002} {\bibfield  {journal} {\bibinfo  {journal} {Physics Reports}\ }\textbf {\bibinfo {volume} {486}},\ \bibinfo {pages} {75} (\bibinfo {year} {2010})}\BibitemShut {NoStop}%
\bibitem [{\citenamefont {Rosvall}\ and\ \citenamefont {Bergstrom}(2008)}]{RosvallBergstrom2008}%
  \BibitemOpen
  \bibfield  {author} {\bibinfo {author} {\bibfnamefont {M.}~\bibnamefont {Rosvall}}\ and\ \bibinfo {author} {\bibfnamefont {C.~T.}\ \bibnamefont {Bergstrom}},\ }\bibfield  {title} {\bibinfo {title} {Maps of random walks on complex networks reveal community structure},\ }\href {https://doi.org/10.1073/pnas.0706851105} {\bibfield  {journal} {\bibinfo  {journal} {Proceedings of the National Academy of Sciences}\ }\textbf {\bibinfo {volume} {105}},\ \bibinfo {pages} {1118} (\bibinfo {year} {2008})}\BibitemShut {NoStop}%
\bibitem [{\citenamefont {Brandes}\ \emph {et~al.}(2008)\citenamefont {Brandes}, \citenamefont {Delling}, \citenamefont {Gaertler}, \citenamefont {Gorke}, \citenamefont {Hoefer}, \citenamefont {Nikoloski},\ and\ \citenamefont {Wagner}}]{Brandes2008}%
  \BibitemOpen
  \bibfield  {author} {\bibinfo {author} {\bibfnamefont {U.}~\bibnamefont {Brandes}}, \bibinfo {author} {\bibfnamefont {D.}~\bibnamefont {Delling}}, \bibinfo {author} {\bibfnamefont {M.}~\bibnamefont {Gaertler}}, \bibinfo {author} {\bibfnamefont {R.}~\bibnamefont {Gorke}}, \bibinfo {author} {\bibfnamefont {M.}~\bibnamefont {Hoefer}}, \bibinfo {author} {\bibfnamefont {Z.}~\bibnamefont {Nikoloski}},\ and\ \bibinfo {author} {\bibfnamefont {D.}~\bibnamefont {Wagner}},\ }\bibfield  {title} {\bibinfo {title} {On modularity clustering},\ }\href {https://doi.org/10.1109/TKDE.2007.190689} {\bibfield  {journal} {\bibinfo  {journal} {IEEE Transactions on Knowledge and Data Engineering}\ }\textbf {\bibinfo {volume} {20}},\ \bibinfo {pages} {172} (\bibinfo {year} {2008})}\BibitemShut {NoStop}%
\bibitem [{\citenamefont {Blondel}\ \emph {et~al.}(2008)\citenamefont {Blondel}, \citenamefont {Guillaume}, \citenamefont {Lambiotte},\ and\ \citenamefont {Lefebvre}}]{Blondel2008}%
  \BibitemOpen
  \bibfield  {author} {\bibinfo {author} {\bibfnamefont {V.}~\bibnamefont {Blondel}}, \bibinfo {author} {\bibfnamefont {J.-L.}\ \bibnamefont {Guillaume}}, \bibinfo {author} {\bibfnamefont {R.}~\bibnamefont {Lambiotte}},\ and\ \bibinfo {author} {\bibfnamefont {E.}~\bibnamefont {Lefebvre}},\ }\bibfield  {title} {\bibinfo {title} {Fast unfolding of communities in large networks},\ }\bibfield  {journal} {\bibinfo  {journal} {Journal of Statistical Mechanics Theory and Experiment}\ }\textbf {\bibinfo {volume} {2008}},\ \href {https://doi.org/10.1088/1742-5468/2008/10/P10008} {10.1088/1742-5468/2008/10/P10008} (\bibinfo {year} {2008})\BibitemShut {NoStop}%
\bibitem [{\citenamefont {Waltman}\ and\ \citenamefont {van Eck}(2013)}]{Waltman2013}%
  \BibitemOpen
  \bibfield  {author} {\bibinfo {author} {\bibfnamefont {L.}~\bibnamefont {Waltman}}\ and\ \bibinfo {author} {\bibfnamefont {N.~J.}\ \bibnamefont {van Eck}},\ }\bibfield  {title} {\bibinfo {title} {A smart local moving algorithm for large-scale modularity-based community detection},\ }\href {https://doi.org/10.1140/epjb/e2013-40829-0} {\bibfield  {journal} {\bibinfo  {journal} {The European Physical Journal B}\ }\textbf {\bibinfo {volume} {86}},\ \bibinfo {pages} {471} (\bibinfo {year} {2013})}\BibitemShut {NoStop}%
\bibitem [{\citenamefont {Edler}\ \emph {et~al.}(2017)\citenamefont {Edler}, \citenamefont {Bohlin},\ and\ \citenamefont {Rosvall}}]{Edler2017}%
  \BibitemOpen
  \bibfield  {author} {\bibinfo {author} {\bibfnamefont {D.}~\bibnamefont {Edler}}, \bibinfo {author} {\bibfnamefont {L.}~\bibnamefont {Bohlin}},\ and\ \bibinfo {author} {\bibfnamefont {M.}~\bibnamefont {Rosvall}},\ }\bibfield  {title} {\bibinfo {title} {Mapping higher-order network flows in memory and multilayer networks with infomap},\ }\bibfield  {journal} {\bibinfo  {journal} {Algorithms}\ }\textbf {\bibinfo {volume} {10}},\ \href {https://doi.org/10.3390/a10040112} {10.3390/a10040112} (\bibinfo {year} {2017})\BibitemShut {NoStop}%
\bibitem [{\citenamefont {Traag}\ \emph {et~al.}(2019)\citenamefont {Traag}, \citenamefont {Waltman},\ and\ \citenamefont {van Eck}}]{Traag2019}%
  \BibitemOpen
  \bibfield  {author} {\bibinfo {author} {\bibfnamefont {V.~A.}\ \bibnamefont {Traag}}, \bibinfo {author} {\bibfnamefont {L.}~\bibnamefont {Waltman}},\ and\ \bibinfo {author} {\bibfnamefont {N.~J.}\ \bibnamefont {van Eck}},\ }\bibfield  {title} {\bibinfo {title} {From louvain to leiden: guaranteeing well-connected communities},\ }\href {https://doi.org/10.1038/s41598-019-41695-z} {\bibfield  {journal} {\bibinfo  {journal} {Scientific Reports}\ }\textbf {\bibinfo {volume} {9}},\ \bibinfo {pages} {5233} (\bibinfo {year} {2019})}\BibitemShut {NoStop}%
\bibitem [{\citenamefont {Newman}(2004)}]{Newman2004}%
  \BibitemOpen
  \bibfield  {author} {\bibinfo {author} {\bibfnamefont {M.~E.~J.}\ \bibnamefont {Newman}},\ }\bibfield  {title} {\bibinfo {title} {Fast algorithm for detecting community structure in networks},\ }\href {https://doi.org/10.1103/PhysRevE.69.066133} {\bibfield  {journal} {\bibinfo  {journal} {Phys. Rev. E}\ }\textbf {\bibinfo {volume} {69}},\ \bibinfo {pages} {066133} (\bibinfo {year} {2004})}\BibitemShut {NoStop}%
\bibitem [{\citenamefont {Rosvall}\ \emph {et~al.}(2009)\citenamefont {Rosvall}, \citenamefont {Axelsson},\ and\ \citenamefont {Bergstrom}}]{Rosvall2009}%
  \BibitemOpen
  \bibfield  {author} {\bibinfo {author} {\bibfnamefont {M.}~\bibnamefont {Rosvall}}, \bibinfo {author} {\bibfnamefont {D.}~\bibnamefont {Axelsson}},\ and\ \bibinfo {author} {\bibfnamefont {C.~T.}\ \bibnamefont {Bergstrom}},\ }\bibfield  {title} {\bibinfo {title} {The map equation},\ }\href {https://doi.org/10.1140/epjst/e2010-01179-1} {\bibfield  {journal} {\bibinfo  {journal} {The European Physical Journal Special Topics}\ }\textbf {\bibinfo {volume} {178}},\ \bibinfo {pages} {13} (\bibinfo {year} {2009})}\BibitemShut {NoStop}%
\bibitem [{\citenamefont {Good}\ \emph {et~al.}(2010)\citenamefont {Good}, \citenamefont {de~Montjoye},\ and\ \citenamefont {Clauset}}]{Good2010}%
  \BibitemOpen
  \bibfield  {author} {\bibinfo {author} {\bibfnamefont {B.~H.}\ \bibnamefont {Good}}, \bibinfo {author} {\bibfnamefont {Y.-A.}\ \bibnamefont {de~Montjoye}},\ and\ \bibinfo {author} {\bibfnamefont {A.}~\bibnamefont {Clauset}},\ }\bibfield  {title} {\bibinfo {title} {Performance of modularity maximization in practical contexts},\ }\href {https://doi.org/10.1103/PhysRevE.81.046106} {\bibfield  {journal} {\bibinfo  {journal} {Phys. Rev. E}\ }\textbf {\bibinfo {volume} {81}},\ \bibinfo {pages} {046106} (\bibinfo {year} {2010})}\BibitemShut {NoStop}%
\bibitem [{\citenamefont {Lancichinetti}\ and\ \citenamefont {Fortunato}(2012)}]{Lancichinetti2012}%
  \BibitemOpen
  \bibfield  {author} {\bibinfo {author} {\bibfnamefont {A.}~\bibnamefont {Lancichinetti}}\ and\ \bibinfo {author} {\bibfnamefont {S.}~\bibnamefont {Fortunato}},\ }\bibfield  {title} {\bibinfo {title} {Consensus clustering in complex networks},\ }\href {https://doi.org/10.1038/srep00336} {\bibfield  {journal} {\bibinfo  {journal} {Scientific Reports}\ }\textbf {\bibinfo {volume} {2}},\ \bibinfo {pages} {336} (\bibinfo {year} {2012})}\BibitemShut {NoStop}%
\bibitem [{\citenamefont {Coscia}\ \emph {et~al.}(2014)\citenamefont {Coscia}, \citenamefont {Rossetti}, \citenamefont {Giannotti},\ and\ \citenamefont {Pedreschi}}]{Coscia2014}%
  \BibitemOpen
  \bibfield  {author} {\bibinfo {author} {\bibfnamefont {M.}~\bibnamefont {Coscia}}, \bibinfo {author} {\bibfnamefont {G.}~\bibnamefont {Rossetti}}, \bibinfo {author} {\bibfnamefont {F.}~\bibnamefont {Giannotti}},\ and\ \bibinfo {author} {\bibfnamefont {D.}~\bibnamefont {Pedreschi}},\ }\bibfield  {title} {\bibinfo {title} {Uncovering hierarchical and overlapping communities with a local-first approach},\ }\bibfield  {journal} {\bibinfo  {journal} {ACM Trans. Knowl. Discov. Data}\ }\textbf {\bibinfo {volume} {9}},\ \href {https://doi.org/10.1145/2629511} {10.1145/2629511} (\bibinfo {year} {2014})\BibitemShut {NoStop}%
\bibitem [{\citenamefont {Rossetti}\ and\ \citenamefont {Cazabet}(2018)}]{Rossetti2018}%
  \BibitemOpen
  \bibfield  {author} {\bibinfo {author} {\bibfnamefont {G.}~\bibnamefont {Rossetti}}\ and\ \bibinfo {author} {\bibfnamefont {R.}~\bibnamefont {Cazabet}},\ }\bibfield  {title} {\bibinfo {title} {Community discovery in dynamic networks: A survey},\ }\bibfield  {journal} {\bibinfo  {journal} {ACM Comput. Surv.}\ }\textbf {\bibinfo {volume} {51}},\ \href {https://doi.org/10.1145/3172867} {10.1145/3172867} (\bibinfo {year} {2018})\BibitemShut {NoStop}%
\bibitem [{\citenamefont {Tandon}\ \emph {et~al.}(2019)\citenamefont {Tandon}, \citenamefont {Albeshri}, \citenamefont {Thayananthan}, \citenamefont {Alhalabi},\ and\ \citenamefont {Fortunato}}]{Tandon2019}%
  \BibitemOpen
  \bibfield  {author} {\bibinfo {author} {\bibfnamefont {A.}~\bibnamefont {Tandon}}, \bibinfo {author} {\bibfnamefont {A.}~\bibnamefont {Albeshri}}, \bibinfo {author} {\bibfnamefont {V.}~\bibnamefont {Thayananthan}}, \bibinfo {author} {\bibfnamefont {W.}~\bibnamefont {Alhalabi}},\ and\ \bibinfo {author} {\bibfnamefont {S.}~\bibnamefont {Fortunato}},\ }\bibfield  {title} {\bibinfo {title} {Fast consensus clustering in complex networks},\ }\href {https://doi.org/10.1103/PhysRevE.99.042301} {\bibfield  {journal} {\bibinfo  {journal} {Phys. Rev. E}\ }\textbf {\bibinfo {volume} {99}},\ \bibinfo {pages} {042301} (\bibinfo {year} {2019})}\BibitemShut {NoStop}%
\bibitem [{\citenamefont {Calatayud}\ \emph {et~al.}(2019)\citenamefont {Calatayud}, \citenamefont {Bernardo-Madrid}, \citenamefont {Neuman}, \citenamefont {Rojas},\ and\ \citenamefont {Rosvall}}]{Calatayud2019}%
  \BibitemOpen
  \bibfield  {author} {\bibinfo {author} {\bibfnamefont {J.}~\bibnamefont {Calatayud}}, \bibinfo {author} {\bibfnamefont {R.}~\bibnamefont {Bernardo-Madrid}}, \bibinfo {author} {\bibfnamefont {M.}~\bibnamefont {Neuman}}, \bibinfo {author} {\bibfnamefont {A.}~\bibnamefont {Rojas}},\ and\ \bibinfo {author} {\bibfnamefont {M.}~\bibnamefont {Rosvall}},\ }\bibfield  {title} {\bibinfo {title} {Exploring the solution landscape enables more reliable network community detection},\ }\href {https://doi.org/10.1103/PhysRevE.100.052308} {\bibfield  {journal} {\bibinfo  {journal} {Physical Review E}\ }\textbf {\bibinfo {volume} {100}},\ \bibinfo {pages} {52308} (\bibinfo {year} {2019})}\BibitemShut {NoStop}%
\bibitem [{\citenamefont {Nathe}\ \emph {et~al.}(2022)\citenamefont {Nathe}, \citenamefont {Gambuzza}, \citenamefont {Frasca},\ and\ \citenamefont {Sorrentino}}]{Nathe2022}%
  \BibitemOpen
  \bibfield  {author} {\bibinfo {author} {\bibfnamefont {C.}~\bibnamefont {Nathe}}, \bibinfo {author} {\bibfnamefont {L.~V.}\ \bibnamefont {Gambuzza}}, \bibinfo {author} {\bibfnamefont {M.}~\bibnamefont {Frasca}},\ and\ \bibinfo {author} {\bibfnamefont {F.}~\bibnamefont {Sorrentino}},\ }\bibfield  {title} {\bibinfo {title} {Looking beyond community structure leads to the discovery of dynamical communities in weighted networks},\ }\href {https://doi.org/10.1038/s41598-022-08214-z} {\bibfield  {journal} {\bibinfo  {journal} {Scientific Reports}\ }\textbf {\bibinfo {volume} {12}},\ \bibinfo {pages} {4524} (\bibinfo {year} {2022})}\BibitemShut {NoStop}%
\bibitem [{\citenamefont {Zachary}(1977)}]{Zachary1977}%
  \BibitemOpen
  \bibfield  {author} {\bibinfo {author} {\bibfnamefont {W.~W.}\ \bibnamefont {Zachary}},\ }\bibfield  {title} {\bibinfo {title} {An information flow model for conflict and fission in small groups},\ }\href {https://doi.org/10.1086/jar.33.4.3629752} {\bibfield  {journal} {\bibinfo  {journal} {Journal of Anthropological Research}\ }\textbf {\bibinfo {volume} {33}},\ \bibinfo {pages} {452} (\bibinfo {year} {1977})}\BibitemShut {NoStop}%
\bibitem [{\citenamefont {Newman}\ and\ \citenamefont {Girvan}(2004)}]{NewmanGirvan2004}%
  \BibitemOpen
  \bibfield  {author} {\bibinfo {author} {\bibfnamefont {M.~E.~J.}\ \bibnamefont {Newman}}\ and\ \bibinfo {author} {\bibfnamefont {M.}~\bibnamefont {Girvan}},\ }\bibfield  {title} {\bibinfo {title} {Finding and evaluating community structure in networks},\ }\href {https://doi.org/10.1103/PhysRevE.69.026113} {\bibfield  {journal} {\bibinfo  {journal} {Phys. Rev. E}\ }\textbf {\bibinfo {volume} {69}},\ \bibinfo {pages} {026113} (\bibinfo {year} {2004})}\BibitemShut {NoStop}%
\bibitem [{\citenamefont {Gfeller}\ \emph {et~al.}(2005)\citenamefont {Gfeller}, \citenamefont {Chappelier},\ and\ \citenamefont {De~Los~Rios}}]{Gfeller2005}%
  \BibitemOpen
  \bibfield  {author} {\bibinfo {author} {\bibfnamefont {D.}~\bibnamefont {Gfeller}}, \bibinfo {author} {\bibfnamefont {J.-C.}\ \bibnamefont {Chappelier}},\ and\ \bibinfo {author} {\bibfnamefont {P.}~\bibnamefont {De~Los~Rios}},\ }\bibfield  {title} {\bibinfo {title} {Finding instabilities in the community structure of complex networks},\ }\href {https://doi.org/10.1103/PhysRevE.72.056135} {\bibfield  {journal} {\bibinfo  {journal} {Phys. Rev. E}\ }\textbf {\bibinfo {volume} {72}},\ \bibinfo {pages} {056135} (\bibinfo {year} {2005})}\BibitemShut {NoStop}%
\bibitem [{\citenamefont {Agarwal}\ and\ \citenamefont {Kempe}(2008)}]{Agarwal2008}%
  \BibitemOpen
  \bibfield  {author} {\bibinfo {author} {\bibfnamefont {G.}~\bibnamefont {Agarwal}}\ and\ \bibinfo {author} {\bibfnamefont {D.}~\bibnamefont {Kempe}},\ }\bibfield  {title} {\bibinfo {title} {Modularity-maximizing graph communities via mathematical programming},\ }\href {https://doi.org/10.1140/epjb/e2008-00425-1} {\bibfield  {journal} {\bibinfo  {journal} {The European Physical Journal B}\ }\textbf {\bibinfo {volume} {66}},\ \bibinfo {pages} {409} (\bibinfo {year} {2008})}\BibitemShut {NoStop}%
\bibitem [{\citenamefont {Karrer}\ \emph {et~al.}(2008)\citenamefont {Karrer}, \citenamefont {Levina},\ and\ \citenamefont {Newman}}]{Karrer2008}%
  \BibitemOpen
  \bibfield  {author} {\bibinfo {author} {\bibfnamefont {B.}~\bibnamefont {Karrer}}, \bibinfo {author} {\bibfnamefont {E.}~\bibnamefont {Levina}},\ and\ \bibinfo {author} {\bibfnamefont {M.~E.~J.}\ \bibnamefont {Newman}},\ }\bibfield  {title} {\bibinfo {title} {Robustness of community structure in networks},\ }\href {https://doi.org/10.1103/PhysRevE.77.046119} {\bibfield  {journal} {\bibinfo  {journal} {Phys. Rev. E}\ }\textbf {\bibinfo {volume} {77}},\ \bibinfo {pages} {046119} (\bibinfo {year} {2008})}\BibitemShut {NoStop}%
\bibitem [{\citenamefont {Cheng}\ and\ \citenamefont {Shen}(2010)}]{Cheng2010}%
  \BibitemOpen
  \bibfield  {author} {\bibinfo {author} {\bibfnamefont {X.-Q.}\ \bibnamefont {Cheng}}\ and\ \bibinfo {author} {\bibfnamefont {H.-W.}\ \bibnamefont {Shen}},\ }\bibfield  {title} {\bibinfo {title} {Uncovering the community structure associated with the diffusion dynamics on networks},\ }\href {https://doi.org/10.1088/1742-5468/2010/04/P04024} {\bibfield  {journal} {\bibinfo  {journal} {Journal of Statistical Mechanics: Theory and Experiment}\ }\textbf {\bibinfo {volume} {2010}},\ \bibinfo {pages} {P04024} (\bibinfo {year} {2010})}\BibitemShut {NoStop}%
\bibitem [{\citenamefont {Peel}\ \emph {et~al.}(2017)\citenamefont {Peel}, \citenamefont {Larremore},\ and\ \citenamefont {Clauset}}]{Peel2017}%
  \BibitemOpen
  \bibfield  {author} {\bibinfo {author} {\bibfnamefont {L.}~\bibnamefont {Peel}}, \bibinfo {author} {\bibfnamefont {D.~B.}\ \bibnamefont {Larremore}},\ and\ \bibinfo {author} {\bibfnamefont {A.}~\bibnamefont {Clauset}},\ }\bibfield  {title} {\bibinfo {title} {The ground truth about metadata and community detection in networks},\ }\href {https://doi.org/10.1126/sciadv.1602548} {\bibfield  {journal} {\bibinfo  {journal} {Science Advances}\ }\textbf {\bibinfo {volume} {3}},\ \bibinfo {pages} {e1602548} (\bibinfo {year} {2017})}\BibitemShut {NoStop}%
\bibitem [{\citenamefont {Newman}(2006)}]{Newman2006}%
  \BibitemOpen
  \bibfield  {author} {\bibinfo {author} {\bibfnamefont {M.~E.~J.}\ \bibnamefont {Newman}},\ }\bibfield  {title} {\bibinfo {title} {Modularity and community structure in networks},\ }\href {https://doi.org/10.1073/pnas.0601602103} {\bibfield  {journal} {\bibinfo  {journal} {Proceedings of the National Academy of Sciences}\ }\textbf {\bibinfo {volume} {103}},\ \bibinfo {pages} {8577} (\bibinfo {year} {2006})}\BibitemShut {NoStop}%
\bibitem [{\citenamefont {Lancichinetti}\ and\ \citenamefont {Fortunato}(2011)}]{LancichinettiFortunato2011}%
  \BibitemOpen
  \bibfield  {author} {\bibinfo {author} {\bibfnamefont {A.}~\bibnamefont {Lancichinetti}}\ and\ \bibinfo {author} {\bibfnamefont {S.}~\bibnamefont {Fortunato}},\ }\bibfield  {title} {\bibinfo {title} {Limits of modularity maximization in community detection},\ }\href {https://doi.org/10.1103/PhysRevE.84.066122} {\bibfield  {journal} {\bibinfo  {journal} {Phys. Rev. E}\ }\textbf {\bibinfo {volume} {84}},\ \bibinfo {pages} {066122} (\bibinfo {year} {2011})}\BibitemShut {NoStop}%
\bibitem [{\citenamefont {Massen}\ and\ \citenamefont {Doye}(2006)}]{Massen2006}%
  \BibitemOpen
  \bibfield  {author} {\bibinfo {author} {\bibfnamefont {C.~P.}\ \bibnamefont {Massen}}\ and\ \bibinfo {author} {\bibfnamefont {J.~P.~K.}\ \bibnamefont {Doye}},\ }\href@noop {} {\bibinfo {title} {Thermodynamics of community structure}} (\bibinfo {year} {2006}),\ \Eprint {https://arxiv.org/abs/cond-mat/0610077} {arXiv:cond-mat/0610077 [cond-mat.stat-mech]} \BibitemShut {NoStop}%
\bibitem [{\citenamefont {Reijnders}\ \emph {et~al.}(2021)\citenamefont {Reijnders}, \citenamefont {van Leeuwen},\ and\ \citenamefont {van Sebille}}]{Reijnders2021}%
  \BibitemOpen
  \bibfield  {author} {\bibinfo {author} {\bibfnamefont {D.}~\bibnamefont {Reijnders}}, \bibinfo {author} {\bibfnamefont {E.~J.}\ \bibnamefont {van Leeuwen}},\ and\ \bibinfo {author} {\bibfnamefont {E.}~\bibnamefont {van Sebille}},\ }\bibfield  {title} {\bibinfo {title} {Ocean surface connectivity in the arctic: Capabilities and caveats of community detection in lagrangian flow networks},\ }\bibfield  {journal} {\bibinfo  {journal} {{Journal of Geophysical Research: Oceans}}\ }\textbf {\bibinfo {volume} {126}},\ \href {https://doi.org/10.1029/2020JC016416} {10.1029/2020JC016416} (\bibinfo {year} {2021}),\ \bibinfo {note} {e2020JC016416 2020JC016416}\BibitemShut {NoStop}%
\bibitem [{\citenamefont {Riolo}\ and\ \citenamefont {Newman}(2020)}]{Riolo2020}%
  \BibitemOpen
  \bibfield  {author} {\bibinfo {author} {\bibfnamefont {M.~A.}\ \bibnamefont {Riolo}}\ and\ \bibinfo {author} {\bibfnamefont {M.~E.~J.}\ \bibnamefont {Newman}},\ }\bibfield  {title} {\bibinfo {title} {Consistency of community structure in complex networks},\ }\href {https://doi.org/10.1103/PhysRevE.101.052306} {\bibfield  {journal} {\bibinfo  {journal} {Phys. Rev. E}\ }\textbf {\bibinfo {volume} {101}},\ \bibinfo {pages} {052306} (\bibinfo {year} {2020})}\BibitemShut {NoStop}%
\bibitem [{\citenamefont {Lee}\ \emph {et~al.}(2021)\citenamefont {Lee}, \citenamefont {Lee}, \citenamefont {Kim},\ and\ \citenamefont {Kim}}]{Lee2021}%
  \BibitemOpen
  \bibfield  {author} {\bibinfo {author} {\bibfnamefont {D.}~\bibnamefont {Lee}}, \bibinfo {author} {\bibfnamefont {S.~H.}\ \bibnamefont {Lee}}, \bibinfo {author} {\bibfnamefont {B.~J.}\ \bibnamefont {Kim}},\ and\ \bibinfo {author} {\bibfnamefont {H.}~\bibnamefont {Kim}},\ }\bibfield  {title} {\bibinfo {title} {Consistency landscape of network communities},\ }\bibfield  {journal} {\bibinfo  {journal} {Physical Review E}\ }\textbf {\bibinfo {volume} {103}},\ \href {https://doi.org/10.1103/physreve.103.052306} {10.1103/physreve.103.052306} (\bibinfo {year} {2021})\BibitemShut {NoStop}%
\bibitem [{\citenamefont {Guimer\`a}\ \emph {et~al.}(2004)\citenamefont {Guimer\`a}, \citenamefont {Sales-Pardo},\ and\ \citenamefont {Amaral}}]{Guimera2004}%
  \BibitemOpen
  \bibfield  {author} {\bibinfo {author} {\bibfnamefont {R.}~\bibnamefont {Guimer\`a}}, \bibinfo {author} {\bibfnamefont {M.}~\bibnamefont {Sales-Pardo}},\ and\ \bibinfo {author} {\bibfnamefont {L.~A.~N.}\ \bibnamefont {Amaral}},\ }\bibfield  {title} {\bibinfo {title} {Modularity from fluctuations in random graphs and complex networks},\ }\href {https://doi.org/10.1103/PhysRevE.70.025101} {\bibfield  {journal} {\bibinfo  {journal} {Phys. Rev. E}\ }\textbf {\bibinfo {volume} {70}},\ \bibinfo {pages} {025101} (\bibinfo {year} {2004})}\BibitemShut {NoStop}%
\bibitem [{\citenamefont {Rosvall}\ and\ \citenamefont {Bergstrom}(2010)}]{RosvallBergstrom2010}%
  \BibitemOpen
  \bibfield  {author} {\bibinfo {author} {\bibfnamefont {M.}~\bibnamefont {Rosvall}}\ and\ \bibinfo {author} {\bibfnamefont {C.~T.}\ \bibnamefont {Bergstrom}},\ }\bibfield  {title} {\bibinfo {title} {Mapping change in large networks},\ }\href {https://doi.org/10.1371/journal.pone.0008694} {\bibfield  {journal} {\bibinfo  {journal} {PLOS ONE}\ }\textbf {\bibinfo {volume} {5}},\ \bibinfo {pages} {e8694} (\bibinfo {year} {2010})}\BibitemShut {NoStop}%
\bibitem [{\citenamefont {Guimerà}\ \emph {et~al.}(2005)\citenamefont {Guimerà}, \citenamefont {Mossa}, \citenamefont {Turtschi},\ and\ \citenamefont {Amaral}}]{Guimera2005}%
  \BibitemOpen
  \bibfield  {author} {\bibinfo {author} {\bibfnamefont {R.}~\bibnamefont {Guimerà}}, \bibinfo {author} {\bibfnamefont {S.}~\bibnamefont {Mossa}}, \bibinfo {author} {\bibfnamefont {A.}~\bibnamefont {Turtschi}},\ and\ \bibinfo {author} {\bibfnamefont {L.~A.~N.}\ \bibnamefont {Amaral}},\ }\bibfield  {title} {\bibinfo {title} {The worldwide air transportation network: Anomalous centrality, community structure, and cities' global roles},\ }\href {https://doi.org/10.1073/pnas.0407994102} {\bibfield  {journal} {\bibinfo  {journal} {Proceedings of the National Academy of Sciences}\ }\textbf {\bibinfo {volume} {102}},\ \bibinfo {pages} {7794} (\bibinfo {year} {2005})}\BibitemShut {NoStop}%
\bibitem [{\citenamefont {Cheung}\ \emph {et~al.}(2020)\citenamefont {Cheung}, \citenamefont {Wong},\ and\ \citenamefont {Zhang}}]{Cheung2020}%
  \BibitemOpen
  \bibfield  {author} {\bibinfo {author} {\bibfnamefont {T.~K.}\ \bibnamefont {Cheung}}, \bibinfo {author} {\bibfnamefont {C.~W.}\ \bibnamefont {Wong}},\ and\ \bibinfo {author} {\bibfnamefont {A.}~\bibnamefont {Zhang}},\ }\bibfield  {title} {\bibinfo {title} {The evolution of aviation network: Global airport connectivity index 2006–2016},\ }\href {https://doi.org/10.1016/j.tre.2019.101826} {\bibfield  {journal} {\bibinfo  {journal} {Transportation Research Part E: Logistics and Transportation Review}\ }\textbf {\bibinfo {volume} {133}},\ \bibinfo {pages} {101826} (\bibinfo {year} {2020})}\BibitemShut {NoStop}%
\bibitem [{\citenamefont {Conway}\ \emph {et~al.}(2005)\citenamefont {Conway}, \citenamefont {Mode}, \citenamefont {Berman}, \citenamefont {Martin},\ and\ \citenamefont {Hill}}]{Conway2005}%
  \BibitemOpen
  \bibfield  {author} {\bibinfo {author} {\bibfnamefont {G.}~\bibnamefont {Conway}}, \bibinfo {author} {\bibfnamefont {N.}~\bibnamefont {Mode}}, \bibinfo {author} {\bibfnamefont {M.}~\bibnamefont {Berman}}, \bibinfo {author} {\bibfnamefont {S.}~\bibnamefont {Martin}},\ and\ \bibinfo {author} {\bibfnamefont {A.}~\bibnamefont {Hill}},\ }\bibfield  {title} {\bibinfo {title} {Flight safety in alaska: Comparing attitudes and practices of high- and low-risk air carriers},\ }\href@noop {} {\bibfield  {journal} {\bibinfo  {journal} {Aviation, space, and environmental medicine}\ }\textbf {\bibinfo {volume} {76}},\ \bibinfo {pages} {52} (\bibinfo {year} {2005})}\BibitemShut {NoStop}%
\bibitem [{\citenamefont {Sun}\ \emph {et~al.}(2022)\citenamefont {Sun}, \citenamefont {Wandelt},\ and\ \citenamefont {Zhang}}]{Sun2022}%
  \BibitemOpen
  \bibfield  {author} {\bibinfo {author} {\bibfnamefont {X.}~\bibnamefont {Sun}}, \bibinfo {author} {\bibfnamefont {S.}~\bibnamefont {Wandelt}},\ and\ \bibinfo {author} {\bibfnamefont {A.}~\bibnamefont {Zhang}},\ }\bibfield  {title} {\bibinfo {title} {{COVID-19} pandemic and air transportation: Summary of recent research, policy consideration and future research directions},\ }\href {https://doi.org/10.1016/j.trip.2022.100718} {\bibfield  {journal} {\bibinfo  {journal} {Transp Res Interdiscip Perspect}\ }\textbf {\bibinfo {volume} {16}},\ \bibinfo {pages} {100718} (\bibinfo {year} {2022})}\BibitemShut {NoStop}%
\bibitem [{\citenamefont {Lex}\ \emph {et~al.}(2014)\citenamefont {Lex}, \citenamefont {Gehlenborg}, \citenamefont {Strobelt}, \citenamefont {Vuillemot},\ and\ \citenamefont {Pfister}}]{Lex2014}%
  \BibitemOpen
  \bibfield  {author} {\bibinfo {author} {\bibfnamefont {A.}~\bibnamefont {Lex}}, \bibinfo {author} {\bibfnamefont {N.}~\bibnamefont {Gehlenborg}}, \bibinfo {author} {\bibfnamefont {H.}~\bibnamefont {Strobelt}}, \bibinfo {author} {\bibfnamefont {R.}~\bibnamefont {Vuillemot}},\ and\ \bibinfo {author} {\bibfnamefont {H.}~\bibnamefont {Pfister}},\ }\bibfield  {title} {\bibinfo {title} {Upset: Visualization of intersecting sets},\ }\href {https://doi.org/10.1109/TVCG.2014.2346248} {\bibfield  {journal} {\bibinfo  {journal} {IEEE Transactions on Visualization and Computer Graphics}\ }\textbf {\bibinfo {volume} {20}},\ \bibinfo {pages} {1983} (\bibinfo {year} {2014})}\BibitemShut {NoStop}%
\bibitem [{\citenamefont {Kim}\ and\ \citenamefont {Lee}(2019)}]{Kim2019}%
  \BibitemOpen
  \bibfield  {author} {\bibinfo {author} {\bibfnamefont {H.}~\bibnamefont {Kim}}\ and\ \bibinfo {author} {\bibfnamefont {S.~H.}\ \bibnamefont {Lee}},\ }\bibfield  {title} {\bibinfo {title} {Relational flexibility of network elements based on inconsistent community detection},\ }\bibfield  {journal} {\bibinfo  {journal} {Physical Review E}\ }\textbf {\bibinfo {volume} {100}},\ \href {https://doi.org/10.1103/physreve.100.022311} {10.1103/physreve.100.022311} (\bibinfo {year} {2019})\BibitemShut {NoStop}%
\bibitem [{\citenamefont {Palla}\ \emph {et~al.}(2005)\citenamefont {Palla}, \citenamefont {Der{\'e}nyi}, \citenamefont {Farkas},\ and\ \citenamefont {Vicsek}}]{Palla2005}%
  \BibitemOpen
  \bibfield  {author} {\bibinfo {author} {\bibfnamefont {G.}~\bibnamefont {Palla}}, \bibinfo {author} {\bibfnamefont {I.}~\bibnamefont {Der{\'e}nyi}}, \bibinfo {author} {\bibfnamefont {I.}~\bibnamefont {Farkas}},\ and\ \bibinfo {author} {\bibfnamefont {T.}~\bibnamefont {Vicsek}},\ }\bibfield  {title} {\bibinfo {title} {Uncovering the overlapping community structure of complex networks in nature and society},\ }\href {https://doi.org/10.1038/nature03607} {\bibfield  {journal} {\bibinfo  {journal} {Nature}\ }\textbf {\bibinfo {volume} {435}},\ \bibinfo {pages} {814} (\bibinfo {year} {2005})}\BibitemShut {NoStop}%
\bibitem [{\citenamefont {Leskovec}\ \emph {et~al.}(2009)\citenamefont {Leskovec}, \citenamefont {Lang}, \citenamefont {Dasgupta},\ and\ \citenamefont {Mahoney}}]{Leskovec2009}%
  \BibitemOpen
  \bibfield  {author} {\bibinfo {author} {\bibfnamefont {J.}~\bibnamefont {Leskovec}}, \bibinfo {author} {\bibfnamefont {K.~J.}\ \bibnamefont {Lang}}, \bibinfo {author} {\bibfnamefont {A.}~\bibnamefont {Dasgupta}},\ and\ \bibinfo {author} {\bibfnamefont {M.~W.}\ \bibnamefont {Mahoney}},\ }\bibfield  {title} {\bibinfo {title} {{Community Structure in Large Networks: Natural Cluster Sizes and the Absence of Large Well-Defined Clusters}},\ }\href@noop {} {\bibfield  {journal} {\bibinfo  {journal} {Internet Mathematics}\ }\textbf {\bibinfo {volume} {6}},\ \bibinfo {pages} {29 } (\bibinfo {year} {2009})}\BibitemShut {NoStop}%
\bibitem [{\citenamefont {Yang}\ and\ \citenamefont {Leskovec}(2014)}]{YangLeskovec2014}%
  \BibitemOpen
  \bibfield  {author} {\bibinfo {author} {\bibfnamefont {J.}~\bibnamefont {Yang}}\ and\ \bibinfo {author} {\bibfnamefont {J.}~\bibnamefont {Leskovec}},\ }\bibfield  {title} {\bibinfo {title} {Structure and overlaps of ground-truth communities in networks},\ }\bibfield  {journal} {\bibinfo  {journal} {ACM Trans. Intell. Syst. Technol.}\ }\textbf {\bibinfo {volume} {5}},\ \href {https://doi.org/10.1145/2594454} {10.1145/2594454} (\bibinfo {year} {2014})\BibitemShut {NoStop}%
\bibitem [{\citenamefont {Jeub}\ \emph {et~al.}(2015)\citenamefont {Jeub}, \citenamefont {Balachandran}, \citenamefont {Porter}, \citenamefont {Mucha},\ and\ \citenamefont {Mahoney}}]{Jeub2015}%
  \BibitemOpen
  \bibfield  {author} {\bibinfo {author} {\bibfnamefont {L.~G.~S.}\ \bibnamefont {Jeub}}, \bibinfo {author} {\bibfnamefont {P.}~\bibnamefont {Balachandran}}, \bibinfo {author} {\bibfnamefont {M.~A.}\ \bibnamefont {Porter}}, \bibinfo {author} {\bibfnamefont {P.~J.}\ \bibnamefont {Mucha}},\ and\ \bibinfo {author} {\bibfnamefont {M.~W.}\ \bibnamefont {Mahoney}},\ }\bibfield  {title} {\bibinfo {title} {Think locally, act locally: Detection of small, medium-sized, and large communities in large networks},\ }\href {https://doi.org/10.1103/PhysRevE.91.012821} {\bibfield  {journal} {\bibinfo  {journal} {Phys. Rev. E}\ }\textbf {\bibinfo {volume} {91}},\ \bibinfo {pages} {012821} (\bibinfo {year} {2015})}\BibitemShut {NoStop}%
\bibitem [{\citenamefont {Ghalmane}\ \emph {et~al.}(2019)\citenamefont {Ghalmane}, \citenamefont {Hassouni},\ and\ \citenamefont {Cherifi}}]{Ghalmane2019}%
  \BibitemOpen
  \bibfield  {author} {\bibinfo {author} {\bibfnamefont {Z.}~\bibnamefont {Ghalmane}}, \bibinfo {author} {\bibfnamefont {M.~E.}\ \bibnamefont {Hassouni}},\ and\ \bibinfo {author} {\bibfnamefont {H.}~\bibnamefont {Cherifi}},\ }\bibfield  {title} {\bibinfo {title} {Immunization of networks with non-overlapping community structure},\ }\href {https://doi.org/10.1007/s13278-019-0591-9} {\bibfield  {journal} {\bibinfo  {journal} {Social Network Analysis and Mining}\ }\textbf {\bibinfo {volume} {9}},\ \bibinfo {pages} {45} (\bibinfo {year} {2019})}\BibitemShut {NoStop}%
\bibitem [{\citenamefont {Lancichinetti}\ \emph {et~al.}(2008)\citenamefont {Lancichinetti}, \citenamefont {Fortunato},\ and\ \citenamefont {Radicchi}}]{LFR2008}%
  \BibitemOpen
  \bibfield  {author} {\bibinfo {author} {\bibfnamefont {A.}~\bibnamefont {Lancichinetti}}, \bibinfo {author} {\bibfnamefont {S.}~\bibnamefont {Fortunato}},\ and\ \bibinfo {author} {\bibfnamefont {F.}~\bibnamefont {Radicchi}},\ }\bibfield  {title} {\bibinfo {title} {Benchmark graphs for testing community detection algorithms},\ }\href {https://doi.org/10.1103/PhysRevE.78.046110} {\bibfield  {journal} {\bibinfo  {journal} {Phys. Rev. E}\ }\textbf {\bibinfo {volume} {78}},\ \bibinfo {pages} {046110} (\bibinfo {year} {2008})}\BibitemShut {NoStop}%
\bibitem [{\citenamefont {Radicchi}\ \emph {et~al.}(2004)\citenamefont {Radicchi}, \citenamefont {Castellano}, \citenamefont {Cecconi}, \citenamefont {Loreto},\ and\ \citenamefont {Parisi}}]{Radicchi2004}%
  \BibitemOpen
  \bibfield  {author} {\bibinfo {author} {\bibfnamefont {F.}~\bibnamefont {Radicchi}}, \bibinfo {author} {\bibfnamefont {C.}~\bibnamefont {Castellano}}, \bibinfo {author} {\bibfnamefont {F.}~\bibnamefont {Cecconi}}, \bibinfo {author} {\bibfnamefont {V.}~\bibnamefont {Loreto}},\ and\ \bibinfo {author} {\bibfnamefont {D.}~\bibnamefont {Parisi}},\ }\bibfield  {title} {\bibinfo {title} {Defining and identifying communities in networks},\ }\href {https://doi.org/10.1073/pnas.0400054101} {\bibfield  {journal} {\bibinfo  {journal} {Proceedings of the National Academy of Sciences}\ }\textbf {\bibinfo {volume} {101}},\ \bibinfo {pages} {2658} (\bibinfo {year} {2004})}\BibitemShut {NoStop}%
\bibitem [{\citenamefont {Yang}\ \emph {et~al.}(2016)\citenamefont {Yang}, \citenamefont {Algesheimer},\ and\ \citenamefont {Tessone}}]{Yang2016}%
  \BibitemOpen
  \bibfield  {author} {\bibinfo {author} {\bibfnamefont {Z.}~\bibnamefont {Yang}}, \bibinfo {author} {\bibfnamefont {R.}~\bibnamefont {Algesheimer}},\ and\ \bibinfo {author} {\bibfnamefont {C.~J.}\ \bibnamefont {Tessone}},\ }\bibfield  {title} {\bibinfo {title} {A comparative analysis of community detection algorithms on artificial networks},\ }\href {https://doi.org/10.1038/srep30750} {\bibfield  {journal} {\bibinfo  {journal} {Scientific Reports}\ }\textbf {\bibinfo {volume} {6}},\ \bibinfo {pages} {30750} (\bibinfo {year} {2016})}\BibitemShut {NoStop}%
\bibitem [{\citenamefont {Tian}\ and\ \citenamefont {Moriano}(2023)}]{Tian2023}%
  \BibitemOpen
  \bibfield  {author} {\bibinfo {author} {\bibfnamefont {M.}~\bibnamefont {Tian}}\ and\ \bibinfo {author} {\bibfnamefont {P.}~\bibnamefont {Moriano}},\ }\bibfield  {title} {\bibinfo {title} {Robustness of community structure under edge addition},\ }\href {https://doi.org/10.1103/PhysRevE.108.054302} {\bibfield  {journal} {\bibinfo  {journal} {Phys. Rev. E}\ }\textbf {\bibinfo {volume} {108}},\ \bibinfo {pages} {054302} (\bibinfo {year} {2023})}\BibitemShut {NoStop}%
\bibitem [{\citenamefont {Efron}(1979)}]{Efron1979}%
  \BibitemOpen
  \bibfield  {author} {\bibinfo {author} {\bibfnamefont {B.}~\bibnamefont {Efron}},\ }\bibfield  {title} {\bibinfo {title} {{Bootstrap Methods: Another Look at the Jackknife}},\ }\href {https://doi.org/10.1214/aos/1176344552} {\bibfield  {journal} {\bibinfo  {journal} {The Annals of Statistics}\ }\textbf {\bibinfo {volume} {7}},\ \bibinfo {pages} {1 } (\bibinfo {year} {1979})}\BibitemShut {NoStop}%
\bibitem [{\citenamefont {Costenbader}\ and\ \citenamefont {Valente}(2003)}]{Costenbader2003}%
  \BibitemOpen
  \bibfield  {author} {\bibinfo {author} {\bibfnamefont {E.}~\bibnamefont {Costenbader}}\ and\ \bibinfo {author} {\bibfnamefont {T.}~\bibnamefont {Valente}},\ }\bibfield  {title} {\bibinfo {title} {The stability of centrality measures when networks are sampled},\ }\href {https://doi.org/10.1016/S0378-8733(03)00012-1} {\bibfield  {journal} {\bibinfo  {journal} {Social Networks}\ }\textbf {\bibinfo {volume} {25}},\ \bibinfo {pages} {283} (\bibinfo {year} {2003})}\BibitemShut {NoStop}%
\bibitem [{\citenamefont {Kwak}\ \emph {et~al.}(2009)\citenamefont {Kwak}, \citenamefont {Eom}, \citenamefont {Choi}, \citenamefont {Jeong},\ and\ \citenamefont {Moon}}]{Kwak2009}%
  \BibitemOpen
  \bibfield  {author} {\bibinfo {author} {\bibfnamefont {H.}~\bibnamefont {Kwak}}, \bibinfo {author} {\bibfnamefont {Y.-H.}\ \bibnamefont {Eom}}, \bibinfo {author} {\bibfnamefont {Y.}~\bibnamefont {Choi}}, \bibinfo {author} {\bibfnamefont {H.}~\bibnamefont {Jeong}},\ and\ \bibinfo {author} {\bibfnamefont {S.}~\bibnamefont {Moon}},\ }\href {https://arxiv.org/abs/0910.1508} {\bibinfo {title} {Consistent community identification in complex networks}} (\bibinfo {year} {2009}),\ \Eprint {https://arxiv.org/abs/0910.1508} {arXiv:0910.1508 [physics.soc-ph]} \BibitemShut {NoStop}%
\bibitem [{\citenamefont {Ser-Giacomi}\ \emph {et~al.}(2015)\citenamefont {Ser-Giacomi}, \citenamefont {Rossi}, \citenamefont {López},\ and\ \citenamefont {Hernández-García}}]{Ser-Giacomi2015}%
  \BibitemOpen
  \bibfield  {author} {\bibinfo {author} {\bibfnamefont {E.}~\bibnamefont {Ser-Giacomi}}, \bibinfo {author} {\bibfnamefont {V.}~\bibnamefont {Rossi}}, \bibinfo {author} {\bibfnamefont {C.}~\bibnamefont {López}},\ and\ \bibinfo {author} {\bibfnamefont {E.}~\bibnamefont {Hernández-García}},\ }\bibfield  {title} {\bibinfo {title} {{Flow networks: A characterization of geophysical fluid transport}},\ }\href {https://doi.org/10.1063/1.4908231} {\bibfield  {journal} {\bibinfo  {journal} {Chaos: An Interdisciplinary Journal of Nonlinear Science}\ }\textbf {\bibinfo {volume} {25}},\ \bibinfo {pages} {036404} (\bibinfo {year} {2015})}\BibitemShut {NoStop}%
\bibitem [{\citenamefont {Schaub}\ \emph {et~al.}(2012{\natexlab{a}})\citenamefont {Schaub}, \citenamefont {Delvenne}, \citenamefont {Yaliraki},\ and\ \citenamefont {Barahona}}]{Schaub2012a}%
  \BibitemOpen
  \bibfield  {author} {\bibinfo {author} {\bibfnamefont {M.~T.}\ \bibnamefont {Schaub}}, \bibinfo {author} {\bibfnamefont {J.-C.}\ \bibnamefont {Delvenne}}, \bibinfo {author} {\bibfnamefont {S.~N.}\ \bibnamefont {Yaliraki}},\ and\ \bibinfo {author} {\bibfnamefont {M.}~\bibnamefont {Barahona}},\ }\bibfield  {title} {\bibinfo {title} {Markov dynamics as a zooming lens for multiscale community detection: Non clique-like communities and the field-of-view limit},\ }\href {https://doi.org/10.1371/journal.pone.0032210} {\bibfield  {journal} {\bibinfo  {journal} {PLOS ONE}\ }\textbf {\bibinfo {volume} {7}},\ \bibinfo {pages} {1} (\bibinfo {year} {2012}{\natexlab{a}})}\BibitemShut {NoStop}%
\bibitem [{\citenamefont {Schaub}\ \emph {et~al.}(2012{\natexlab{b}})\citenamefont {Schaub}, \citenamefont {Lambiotte},\ and\ \citenamefont {Barahona}}]{Schaub2012b}%
  \BibitemOpen
  \bibfield  {author} {\bibinfo {author} {\bibfnamefont {M.~T.}\ \bibnamefont {Schaub}}, \bibinfo {author} {\bibfnamefont {R.}~\bibnamefont {Lambiotte}},\ and\ \bibinfo {author} {\bibfnamefont {M.}~\bibnamefont {Barahona}},\ }\bibfield  {title} {\bibinfo {title} {Encoding dynamics for multiscale community detection: Markov time sweeping for the map equation},\ }\href {https://doi.org/10.1103/PhysRevE.86.026112} {\bibfield  {journal} {\bibinfo  {journal} {Phys. Rev. E Stat. Nonlin. Soft Matter Phys.}\ }\textbf {\bibinfo {volume} {86}},\ \bibinfo {pages} {026112} (\bibinfo {year} {2012}{\natexlab{b}})}\BibitemShut {NoStop}%
\bibitem [{\citenamefont {Edler}\ \emph {et~al.}(2022)\citenamefont {Edler}, \citenamefont {Smiljanić}, \citenamefont {Holmgren}, \citenamefont {Antonelli},\ and\ \citenamefont {Rosvall}}]{Edler2022}%
  \BibitemOpen
  \bibfield  {author} {\bibinfo {author} {\bibfnamefont {D.}~\bibnamefont {Edler}}, \bibinfo {author} {\bibfnamefont {J.}~\bibnamefont {Smiljanić}}, \bibinfo {author} {\bibfnamefont {A.}~\bibnamefont {Holmgren}}, \bibinfo {author} {\bibfnamefont {A.}~\bibnamefont {Antonelli}},\ and\ \bibinfo {author} {\bibfnamefont {M.}~\bibnamefont {Rosvall}},\ }\href@noop {} {\bibinfo {title} {Variable markov dynamics as a multi-focal lens to map multi-scale complex networks}} (\bibinfo {year} {2022}),\ \Eprint {https://arxiv.org/abs/2211.04287} {arXiv:2211.04287 [physics.soc-ph]} \BibitemShut {NoStop}%
\bibitem [{\citenamefont {Clauset}\ \emph {et~al.}(2004)\citenamefont {Clauset}, \citenamefont {Newman},\ and\ \citenamefont {Moore}}]{CNM2004}%
  \BibitemOpen
  \bibfield  {author} {\bibinfo {author} {\bibfnamefont {A.}~\bibnamefont {Clauset}}, \bibinfo {author} {\bibfnamefont {M.~E.~J.}\ \bibnamefont {Newman}},\ and\ \bibinfo {author} {\bibfnamefont {C.}~\bibnamefont {Moore}},\ }\bibfield  {title} {\bibinfo {title} {Finding community structure in very large networks},\ }\href {https://doi.org/10.1103/PhysRevE.70.066111} {\bibfield  {journal} {\bibinfo  {journal} {Phys. Rev. E}\ }\textbf {\bibinfo {volume} {70}},\ \bibinfo {pages} {066111} (\bibinfo {year} {2004})}\BibitemShut {NoStop}%
\bibitem [{\citenamefont {Kirkpatrick}\ \emph {et~al.}(1983)\citenamefont {Kirkpatrick}, \citenamefont {Gelatt},\ and\ \citenamefont {Vecchi}}]{Kirkpatrick1983}%
  \BibitemOpen
  \bibfield  {author} {\bibinfo {author} {\bibfnamefont {S.}~\bibnamefont {Kirkpatrick}}, \bibinfo {author} {\bibfnamefont {C.~D.}\ \bibnamefont {Gelatt}},\ and\ \bibinfo {author} {\bibfnamefont {M.~P.}\ \bibnamefont {Vecchi}},\ }\bibfield  {title} {\bibinfo {title} {Optimization by simulated annealing},\ }\href {https://doi.org/10.1126/science.220.4598.671} {\bibfield  {journal} {\bibinfo  {journal} {Science}\ }\textbf {\bibinfo {volume} {220}},\ \bibinfo {pages} {671} (\bibinfo {year} {1983})}\BibitemShut {NoStop}%
\bibitem [{\citenamefont {Metropolis}\ \emph {et~al.}(1953)\citenamefont {Metropolis}, \citenamefont {Rosenbluth}, \citenamefont {Rosenbluth}, \citenamefont {Teller},\ and\ \citenamefont {Teller}}]{Metropolis1953}%
  \BibitemOpen
  \bibfield  {author} {\bibinfo {author} {\bibfnamefont {N.}~\bibnamefont {Metropolis}}, \bibinfo {author} {\bibfnamefont {A.~W.}\ \bibnamefont {Rosenbluth}}, \bibinfo {author} {\bibfnamefont {M.~N.}\ \bibnamefont {Rosenbluth}}, \bibinfo {author} {\bibfnamefont {A.~H.}\ \bibnamefont {Teller}},\ and\ \bibinfo {author} {\bibfnamefont {E.}~\bibnamefont {Teller}},\ }\bibfield  {title} {\bibinfo {title} {{Equation of State Calculations by Fast Computing Machines}},\ }\href {https://doi.org/10.1063/1.1699114} {\bibfield  {journal} {\bibinfo  {journal} {The Journal of Chemical Physics}\ }\textbf {\bibinfo {volume} {21}},\ \bibinfo {pages} {1087} (\bibinfo {year} {1953})}\BibitemShut {NoStop}%
\bibitem [{\citenamefont {Hastings}(1970)}]{Hastings1970}%
  \BibitemOpen
  \bibfield  {author} {\bibinfo {author} {\bibfnamefont {W.~K.}\ \bibnamefont {Hastings}},\ }\bibfield  {title} {\bibinfo {title} {Monte carlo sampling methods using markov chains and their applications},\ }\href {http://www.jstor.org/stable/2334940} {\bibfield  {journal} {\bibinfo  {journal} {Biometrika}\ }\textbf {\bibinfo {volume} {57}},\ \bibinfo {pages} {97} (\bibinfo {year} {1970})}\BibitemShut {NoStop}%
\end{thebibliography}%

\end{document}